\documentclass[10pt,journal]{IEEEtran}
\usepackage[T1]{fontenc}
\usepackage[utf8]{inputenc}
\usepackage{amsmath,amssymb,graphicx,booktabs,array,tabularx}
\usepackage{cite}
\usepackage{multirow}
\usepackage{xcolor}
\usepackage{float}
\definecolor{RQBlue}{HTML}{245E7A}
\definecolor{RQBorder}{HTML}{ADC4D0}
\definecolor{RQBackground}{HTML}{F0F5F8}
\floatstyle{plain}
\newfloat{rqsummary}{htb}{lorq}
\newcommand{\RQTakeaway}[3][!htb]{%
  \begingroup
  \setlength{\intextsep}{5pt plus 1pt minus 1pt}%
  \begin{rqsummary}[#1]
  \centering
  \setlength{\fboxsep}{4.5pt}%
  \setlength{\fboxrule}{0.45pt}%
  \fcolorbox{RQBorder}{RQBackground}{%
    \begin{minipage}{\dimexpr\linewidth-2\fboxsep-2\fboxrule\relax}
      \normalfont\normalsize\raggedright\setlength{\parindent}{0pt}%
      {\color{RQBlue}\bfseries #2\par}%
      \vspace{3pt}%
      #3%
    \end{minipage}%
  }%
  \end{rqsummary}%
  \endgroup
}

\usepackage{colortbl}
\definecolor{ProofInk}{HTML}{224B69}
\definecolor{ProofBlueWash}{HTML}{EFF4F6}
\definecolor{ProofAmberWash}{HTML}{FBF2E9}
\definecolor{ProofWarmInk}{HTML}{82441F}
\definecolor{ProofRule}{HTML}{A9B7BE}
\newcommand{\ProofHead}[1]{{\color{ProofInk}\bfseries\boldmath #1}}
\newcommand{\ProofPair}[1]{\textcolor{ProofWarmInk}{\textbf{#1}}}
\newcommand{\ProofKeep}[1]{\textcolor{ProofInk}{\textbf{#1}}}

\newsavebox{\ProofMeasured}
\newcommand{\ProofBand}[6]{%
  \setbox\ProofMeasured=\vbox{\kern#4\hbox{#6}\kern#5}%
  \hbox{\rlap{\hspace*{#1}{\color{#3}\vrule width #2 height \ht\ProofMeasured depth \dp\ProofMeasured}}\box\ProofMeasured}}
\newcommand{\ProofRuleLine}[2]{\hbox{\color{#1}\vrule width \textwidth height #2 depth 0pt}}

\usepackage[hidelinks,breaklinks=true]{hyperref}
\renewcommand{\thesection}{\arabic{section}}
\renewcommand{\thesubsection}{\thesection.\arabic{subsection}}
\ifcsname thesubsectiondis\endcsname
\renewcommand{\thesubsectiondis}{\thesubsection.}
\fi
\renewcommand{\thetable}{\arabic{table}}
\renewcommand{\thefigure}{\arabic{figure}}
\graphicspath{{figures/}{appendix/}}

\title{Slow Cleanup, Fast Extraction: Persistent Permissions and Fund Activity during Recovery from Ethereum Authorization Incidents}
\author{Zhang~Sheng\textsuperscript{*}, Yiran~Huang\textsuperscript{*},
Shen~Wang, Shengchen~Duan,\\
Shengrong~Zhu, Benrui~Chen, Qinlong~Qiu, Dan~Wang,
and Daoyuan~Wu\textsuperscript{\ensuremath{\dagger}}%
\thanks{\textsuperscript{*}Zhang Sheng and Yiran Huang contributed equally to this work.}%
\thanks{\textsuperscript{\ensuremath{\dagger}}Corresponding author: Daoyuan Wu.}%
\thanks{Zhang Sheng and Yiran Huang are with the School of Information Engineering, Shaoxing Vocational \& Technical College, Shaoxing, China.}%
\thanks{Shengrong Zhu, Benrui Chen, Qinlong Qiu, and Dan Wang are with Shaoxing Vocational \& Technical College, Shaoxing, China.}%
\thanks{Shen Wang is with City University of Hong Kong, Hong Kong, China (e-mail: \href{mailto:shenwang918@gmail.com}{\nolinkurl{shenwang918@gmail.com}}).}%
\thanks{Shengchen Duan is with Singapore Management University, Singapore (e-mail: \href{mailto:sc.duan.2024@phdcs.smu.edu.sg}{\nolinkurl{sc.duan.2024@phdcs.smu.edu.sg}}).}%
\thanks{Daoyuan Wu is with Lingnan University, Hong Kong SAR, China (e-mail: \href{mailto:daoyuanwu@ln.edu.hk}{\nolinkurl{daoyuanwu@ln.edu.hk}}; ORCID: \href{https://orcid.org/0000-0002-3752-0718}{0000-0002-3752-0718}).}}

\date{}
\begin{document}
\maketitle
\begin{abstract}
Ethereum users authorize applications to transfer tokens on their behalf. These approvals can expose later deposits when application vulnerabilities are exploited. We investigate when remediation justifies ending emergency monitoring of existing approvals. We track 403 approval relationships across six incidents for 30 days after disclosure and validate the analysis on two additional protocols. Linking remediation, approval histories, and transaction order distinguishes spending under retained permissions from spending under later allowance settings. In Unibot, half the sampled approvals were explicitly cleared by approximately 5 hours and 35 minutes after disclosure. Before allowance clearing, two repurchases by one wallet were followed by repeat extractions after 60 and 84 seconds, respectively. In Multichain, incoming funds were transferred out using an existing approval within one block (a batch of transactions). The account's token balance was zero at both block boundaries. Retrospectively, verified entry-point closure evidence reduced decisions to retain approvals across four checkpoints from 1,023 to 176. With only evidence confirmed available at the decision time, the count remained 1,023. These findings support continuing, ending, or reassessing monitoring of the original incident path. Our open-source toolkit reconstructs permission histories and fund activity and reproduces monitoring decisions, helping responders identify permissions requiring attention.
\end{abstract}

\section{Introduction}\label{sec:1}

Ethereum applications typically require users to grant permission to transfer a particular token on their behalf \cite{ref01}. This approval can remain valid after an individual transaction ends and support repeated transfers. If an application's transfer logic can be abused, attackers may use retained approvals to transfer users' assets through that application \cite{ref02,ref03}. Even after an account has been emptied, newly arriving funds can still be transferred through the retained approval. Recovery therefore concerns both the software issue and the transfer permissions remaining on accounts.

Project teams can patch software or disable entry points, while account holders can revoke an application's permission to spend a token. Public response materials include service suspension announcements \cite{ref04} and notices distinguishing internal repairs from user revocations \cite{ref05,ref06}. These measures may reach completion at different times during incident recovery. An application repaired at its original address may continue using existing approvals for normal service. An account holder may grant a new approval after revocation. Recovery analysis must therefore connect software controls, permission changes, and subsequent use.

A remediation gap remains when a measure has been implemented but the conditions enabling the original transfers still hold. Suspending a website or service does not automatically revoke on-chain approvals. If the affected call remains usable, retained permissions may still enable transfers. A zero balance means that the account currently has none of that token to transfer; later deposits may still be extracted \cite{ref03}. Ending monitoring on the basis of an announcement or a current zero balance can therefore remove subsequent permission use from observation. For each approval relationship, we check whether the affected call has been blocked and whether its existing permission has been removed.

Prior research quantifies spendable amounts from allowances and balances, examines future-deposit risks, and reconstructs approval--use sequences \cite{ref03}. OSPREY analyzes controllable transferFrom vulnerabilities and historically exploitable instances \cite{ref02}. Other studies examine emergency pauses and rescue windows \cite{ref07}, phishing victims' remedial actions \cite{ref08}, and repeated extraction and response timing \cite{ref09}. Closeout decisions additionally require aligning these observations with remediation scope and effective time. Subsequent spending may use a pre-disclosure permission or a newly granted approval. We examine the supporting permission separately from the entry point used and the controls in effect. Distinguishing these aspects requires both cleanup measurements and transaction-level permission histories.

This paper addresses two research questions. \textbf{RQ1: How do existing approval relationships and the call paths responsible for an incident change after public disclosure?} \textbf{RQ2: How do permissions persist or become reestablished after an incident, and which approvals support subsequent spending?} RQ1 establishes changes in the affected path and permission state; RQ2 identifies the permissions supporting subsequent operations. Connecting both with returning funds supports decisions to continue emergency monitoring, exit the original incident path, or reassess after a later change.

We align public remediation records, historical allowances and balances, approval operations, and transaction execution evidence by relationship and time. The study follows 403 approval relationships across six public incidents for 30 days after disclosure and examines 170 candidate approval-related logs. We reconstruct 187 replenishment episodes, from a balance becoming positive until it returns to zero or observation ends. After freezing the analysis rules, we conduct external protocol validation on Dexible and Multichain under different remediation conditions.

\begin{figure*}[tp]
\centering
\includegraphics[width=\textwidth,height=0.82\textheight,keepaspectratio]{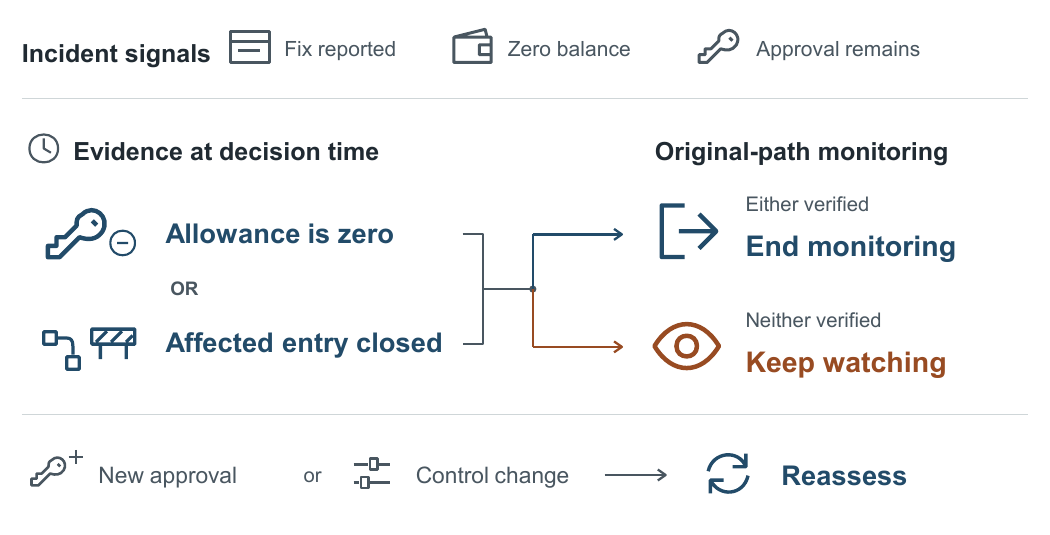}
\renewcommand{\thefigure}{1}
\caption{When emergency monitoring can end for an existing approval. A verified zero allowance or evidence that the affected entry point is closed supports exit from original-path monitoring. A new approval or a change in control calls for reassessment.}\label{fig:1}
\end{figure*}

In Unibot, half of 44 initially positive allowances were first explicitly set to zero about 5 hours and 35 minutes after disclosure. Before its allowance was cleared, one wallet bought tokens back twice and experienced renewed extraction after 60 seconds and 84 seconds, respectively. Neither interval contained a new positive allowance setting. We cross-check these episodes against contemporaneous public alerts \cite{ref10}. One external Multichain relationship shows a deposit and use of a pre-disclosure allowance within the same block, with zero balances before and after. Its transfer remains classified as third-party-directed, with purpose unknown. Conversely, all 97 design-set relationships with positive allowances and balances on day 30 had evidence of original-path closure. Figure~\ref{fig:1} connects these observations to relationship-level closeout decisions.

The paper makes three contributions:

\begin{itemize}
\item \textbf{Approval-specific closeout criteria.} We separately verify permission removal and affected-entry-point closure to identify grounds for ending emergency monitoring. Renewed approval or changed entry-point state requires reassessment.
\item \textbf{Fund-activity evidence for recovery decisions.} We connect cleanup progress to repeated extraction, same-block spending, and continued use after repair. These histories explain why zero balances can coexist with continuing exposure and retained approvals with grounds for exit.
\item \textbf{Recomputable monitoring decisions and an open-source toolkit.} Across four checkpoints in six cases, verified retrospective closure evidence reduces retention decisions from 1,023 to 176. Using only evidence confirmed as available at the time leaves 1,023. These results motivate recovery records specifying controlled entry points, scope, and effective time. The \href{https://github.com/dcszhang/Authorization-Recovery-Toolkit}{Authorization Recovery Toolkit} identifies approval operations, reconstructs fund activity, and recomputes monitoring lists with traceable decision grounds.
\end{itemize}

\section{Background and Recovery Model}\label{sec:2}

\subsection{How Approvals Work}\label{sec:2.1}

An approval permits an application to transfer tokens without depositing them into the application. The token contract separately records account balances and allowances \cite{ref01}. An approval relationship comprises chain, owner, token, and spender: the network, token-holding account, token contract, and address receiving transfer permission. We call the selected approved contract the spender under study.

For relationship r at block b, A denotes remaining allowance, B denotes balance, and nominal spending capacity is \(C=\min(A,B)\) \cite{ref03}. An allowance of 100 tokens with a balance of 20 therefore gives capacity 20. If B becomes zero while A remains positive, capacity temporarily vanishes but permission persists. Later deposits may again become transferable, provided the corresponding software call is executable.

Accounts set or overwrite allowances through operations such as approve; authorized spenders request transfers through transferFrom \cite{ref01}. An explicit zero setting removes transfer eligibility, while spending can also consume allowance. Allowance changes and approval-log semantics depend on the historical token implementation.

\subsection{Approval Risks and Two Types of Remediation}\label{sec:2.2}

An approval vulnerability lets an external caller bypass restrictions and cause an authorized application to make unintended transfers \cite{ref02}. The original incident path comprises the entry points and call relationships verified as relevant to the incident. Control evidence applies to this scope; fund use also requires valid permission and transferable assets.

Software-side remediation patches logic or disables affected entry points; account-side remediation changes a spender's permission for a particular owner and token. The former can control a shared entry point, whereas the latter changes an individual relationship. After repair at the same address, normal business may continue using retained approvals. Permissions no longer needed require individual cancellation. Appendix~\ref{app:A.4}, Figure~\ref{fig:A2}, illustrates these distinct control targets and explains proxy entry points and implementations \cite{ref11,ref12}.

\subsection{Basic Units of Observation during Recovery}\label{sec:2.3}

We link public disclosure, software controls, permission changes, and fund operations through the same relationship, retaining each observation's time and target. A block contains sequential transactions; its final state records only their combined outcome. Transaction and log indices establish the order within a block.

The first explicit zero setting is a cleanup operation; a current A=0 may instead result from consumption. We record each verified positive allowance-setting operation separately in the permission history. We attribute spending to a pre-disclosure allowance only when that allowance was positive before disclosure and no positive allowance-setting operation intervened. Allowance reductions caused by spending remain consumption of the existing permission. Spending supported by later settings is recorded separately, distinguishing repeated calls to one contract from continued use of one permission.

A replenishment episode begins when B changes from zero to positive after a transaction and ends when a later transaction returns B to zero. If the balance remains positive at the end of observation, the episode is unfinished, or statistically right-censored. Deposit-to-spending intervals are calculated separately from the duration of an episode. An entire episode can occur between adjacent transactions within one block, leaving no positive end-of-block balance. Relationships, episodes, transactions, and logs are counted separately because one relationship can contain multiple episodes and one transaction can produce multiple logs.

\section{Study Design and Evidence Reconstruction}\label{sec:3}

We track permission and fund changes through 403 fixed approval relationships and separately sample 170 candidate post-incident positive-allowance logs. The event sample estimates the composition and timing of approval operations. The 187 replenishment episodes derive from complete fixed-relationship histories. These samples retain separate denominators, as do the external cases. Evidence is linked by chain, owner, token, and spender, with states bound to exact blocks and hashes and operations ordered by transaction and log indices.

\subsection{Cases, Samples, and Observation Windows}\label{sec:3.1}

The six-case design set covers shared-path closure, repair at the original address, service suspension, and relationship-level cleanup. Pilot research and eligibility screening verified each spender, selected token, and historical evidence before sampling. After establishing and freezing the analytical rules, we selected Dexible/TRU and Multichain/WETH for external validation (Table~\ref{tab:1}). Results concern each case's selected token and spender.

\begin{table*}[!tp]
\centering\footnotesize
\renewcommand{\thetable}{1}
\caption{Cases and sampling frames.}\label{tab:1}
\setlength{\tabcolsep}{4pt}\renewcommand{\arraystretch}{1.18}
\vbox{\offinterlineskip
\ProofRuleLine{ProofInk}{0.8pt}
\ProofBand{0pt}{\textwidth}{ProofBlueWash}{2.5pt}{1.6pt}{\begin{tabular}{>{\raggedright\arraybackslash}p{\dimexpr 0.11000\textwidth-2\tabcolsep\relax}>{\raggedright\arraybackslash}p{\dimexpr 0.12000\textwidth-2\tabcolsep\relax}>{\raggedright\arraybackslash}p{\dimexpr 0.08000\textwidth-2\tabcolsep\relax}>{\raggedright\arraybackslash}p{\dimexpr 0.35000\textwidth-2\tabcolsep\relax}>{\raggedleft\arraybackslash}p{\dimexpr 0.17000\textwidth-2\tabcolsep\relax}>{\raggedleft\arraybackslash}p{\dimexpr 0.17000\textwidth-2\tabcolsep\relax}}
\ProofHead{Study group} & \ProofHead{Case} & \ProofHead{Token} & \ProofHead{Remediation target} & \ProofHead{\shortstack[r]{Relationships\\\(N / n\)}} & \ProofHead{\shortstack[r]{Log candidates\\\(N / n\)}} \\
\end{tabular}}
\ProofRuleLine{ProofRule}{0.35pt}
\ProofBand{0pt}{\textwidth}{white}{1.5pt}{1.5pt}{\begin{tabular}{>{\raggedright\arraybackslash}p{\dimexpr 0.11000\textwidth-2\tabcolsep\relax}>{\raggedright\arraybackslash}p{\dimexpr 0.12000\textwidth-2\tabcolsep\relax}>{\raggedright\arraybackslash}p{\dimexpr 0.08000\textwidth-2\tabcolsep\relax}>{\raggedright\arraybackslash}p{\dimexpr 0.35000\textwidth-2\tabcolsep\relax}>{\raggedleft\arraybackslash}p{\dimexpr 0.17000\textwidth-2\tabcolsep\relax}>{\raggedleft\arraybackslash}p{\dimexpr 0.17000\textwidth-2\tabcolsep\relax}}
\multirow[t]{6}{\linewidth}{\raggedright\textcolor{ProofInk}{\textbf{Design set}}} & Socket & USDC & Disable the old route & 9,298 / 100 & 1,754 / 50 \\
 & LiFi & USDC & Remove the selector & 51,250 / 100 & 8,323 / 50 \\
 & Maestro & MOG & Upgrade the proxy & 47 / 47 & 59 / 50 \\
 & Unibot & UNIBOT & Suspend the service and clean up individual relationships & 44 / 44 & 20 / 20 \\
 & Dolomite & USDC & Remove entry-point eligibility and pause the proxy & 2,977 / 100 & 0 / 0 \\
 & Sushi RP2 & WETH & Interface remediation and revocation notice & 12 / 12 & 0 / 0 \\
\end{tabular}}
\ProofRuleLine{ProofRule}{0.35pt}
\ProofBand{0pt}{\textwidth}{white}{1.5pt}{1.5pt}{\begin{tabular}{>{\raggedright\arraybackslash}p{\dimexpr 0.11000\textwidth-2\tabcolsep\relax}>{\raggedright\arraybackslash}p{\dimexpr 0.12000\textwidth-2\tabcolsep\relax}>{\raggedright\arraybackslash}p{\dimexpr 0.08000\textwidth-2\tabcolsep\relax}>{\raggedright\arraybackslash}p{\dimexpr 0.35000\textwidth-2\tabcolsep\relax}>{\raggedleft\arraybackslash}p{\dimexpr 0.17000\textwidth-2\tabcolsep\relax}>{\raggedleft\arraybackslash}p{\dimexpr 0.17000\textwidth-2\tabcolsep\relax}}
\multirow[t]{2}{\linewidth}{\raggedright\textcolor{ProofInk}{\textbf{External validation}}} & Dexible & TRU & Pause the old entry point & 1 / 1 & --- \\
 & Multichain & WETH & Repair the pool and issue an old-router revocation notice & 4,573 / 20 & --- \\
\end{tabular}}
\ProofRuleLine{ProofInk}{0.65pt}
}
\par\smallskip
\begin{minipage}{\textwidth}\footnotesize\raggedright
N: complete frame; n: frozen sample. External cases use fixed relationship histories, without separate population event sampling (shown as a dash).
\end{minipage}
\end{table*}

The relationship frame contains positive-allowance Approval logs from spender deployment to disclosure, deduplicated by relationship key. Relationships whose last positive log falls within the preceding 90 days form the recent stratum; the rest form the historical stratum. For tokens emitting Approval during spending, recency denotes log activity. The selected list remains fixed throughout observation.

RQ2 uses a separately constructed 30-day positive-log frame because post-incident operations can involve accounts outside that list. Candidates are stratified into 0--1, 1--7, and 7--30 days. Zero settings are reconstructed separately from complete fixed-relationship histories. Each design case includes at most 100 relationships and 50 candidates; external cases include at most 20 relationships. Small frames are included fully and larger frames use stratified quotas. A prespecified seed and hash ordering determine all lists, which are frozen before outcome queries. Appendix~\ref{app:C.1} gives quotas, the seed, and weights.

Researchers had prior exposure to both external protocols through literature prescreening. The sole Dexible relationship uses the account queried during technical eligibility tests; none of the 20 Multichain relationships overlaps those test accounts. Protocol exposure, account exposure, and sample freezing are recorded separately.

For time-based checkpoints, we read A and B at the latest block at or before the target timestamp, retaining adjacent blocks and hashes for verification. We observe pre-disclosure, disclosure, first identifiable remediation, D1, D7, and D30 states. Repeated same-block records are deduplicated, with their checkpoint labels, block numbers, and relative times preserved. Sushi has no remediation checkpoint because its interface-remediation time is unknown. Unibot uses a block near the time of its suspension announcement. Actual on-chain control blocks are recorded separately.

The design set uses the lower bound of each disclosure timestamp's second- or millisecond-resolution unit. Its pre-disclosure state is taken from the block immediately preceding the block corresponding to that bound. External cases use the disclosure upper bound. Dexible has 1-millisecond precision; Multichain's interval spans 74 hours, from 2022-01-16 10:00 to 2022-01-19 12:00 UTC. Multichain's main analysis starts at the upper bound, reports within-interval activity separately, and includes a 30-day sensitivity analysis from the lower bound. External initial positive-allowance groups use frozen states before the lower bound. Multichain retains all 19 initially positive relationships in its denominator, including the two cleared before the upper bound. At that bound, 17 relationships still have positive allowances. Appendix~\ref{app:A} records time precision and anchors.

\subsection{Reconstructing Permissions and Fund Activity}\label{sec:3.2}

Token storage, interfaces, and events can disagree \cite{ref13}, so approval logs require implementation-specific interpretation. UNIBOT and TRU can emit Approval for allowance remaining after spending. We check each candidate's successful receipt, log, and call arguments, and identify the token implementation active at the relevant block. We classify it as a positive setting, consumption of an existing allowance, or unknown; explicit zero settings are separate cleanup operations.

Reconstruction begins with each relationship's initial state and complete event window, applies effects in execution order, and checks historical state anchors. This connects individual operations to the observed allowance and balance at each checkpoint. Allowances that cannot be reconstructed at transaction boundaries remain UNKNOWN, while verified state anchors are retained separately. Appendix~\ref{app:B.1} preserves the TRU, USDC, WETH, and MOG reconstruction rules.

For complex transactions, we identify the actual token caller from the complete execution trace after excluding reverted calls and all their descendants. A pre-disclosure allowance supports spending only if it was positive before disclosure and no subsequent positive allowance-setting operation preceded that spending. A positive setting after zero supports a subsequent permission; explicitly resetting a positive allowance also constitutes a new setting. Consumption continues the existing permission, and same-block changes follow execution order. Increases, repeated settings, reductions, and reapproval after zero require sufficiently established prior states.

Initiators and destinations distinguish account-initiated from third-party-directed activity. Normal business, attacks, and rescues require separate purpose evidence, so confirmed calls, permission sources, and undetermined purposes remain distinguishable.

\subsection{Remediation Evidence, Measures, and Verification}\label{sec:3.3}

We map incident-time implementations, configuration changes, and historical block states to affected entry points. This establishes each control's scope, effective time, and covered approval relationships. Entry points are classified as verified closed, still in their original configuration, or undetermined. Undetermined means that available evidence cannot establish whether the entry point is open or closed. Configuration and actual spending remain separate observations. Where event records fully cover the relevant changes, we check them through D30; otherwise, we report only verified points. We distinguish the suspension announcement, effective on-chain closure, and availability of sufficient closure evidence. These times are recorded separately, and unavailable times remain explicitly unknown. Scoped closure evidence supports exit from the corresponding original path.

For RQ1, the initial positive-allowance group remains fixed while we measure the weighted cumulative proportion receiving a first explicit zero setting. The times reaching 50\% and 90\% are \(t_{50}\) and \(t_{90}\), indicating when half and nine-tenths of these permissions have been explicitly cleared. Failure to reach a threshold differs from an undetermined zero-setting time. We separately report consumption to zero, current allowance, first explicit zero setting, balance, and affected-path state.

For RQ2, we compare positive settings, consumption, and unknown operations across the three event-sampling periods. Fixed-relationship zero settings and subsequent settings remain separate history-based observations. For each replenishment episode, we record arrival-time allowance and path state, subsequent application spending, its permission source, and deposit-to-spending order and interval. Relationship and episode counts remain separate because funds can return repeatedly.

Sampling-fraction weights determine how many frame objects each sampled item represents. We retain observed counts alongside estimated counts, proportions, and uncertainty intervals. UNKNOWN records remain in their original denominators and contribute to interval calculations. Appendix~\ref{app:C.1} specifies weighting, interval assumptions, and both Multichain timing conventions; Appendix~\ref{app:C.2} specifies allowance storage, conversion, and magnitude thresholds.

Each state record pair contains the allowance and balance of one approval relationship read at the same historical block. All 2,406 pairs in the design set are complete; 470 pairs were checked against independent sources. Appendix~\ref{app:A} supplies external-case execution evidence, transition checks, raw responses, and source locations for item-by-item review.

\subsection{Implementation and Validation}\label{sec:3.4}

The open-source \href{https://github.com/dcszhang/Authorization-Recovery-Toolkit}{Authorization Recovery Toolkit} reconstructs permission and fund activity and aggregates reviewed records into cleanup measures, end states, episode groups, and monitoring lists. Its transaction component identifies allowance settings and spending, reconstructs states, and checks whether the authorized application actually used the account's permission. Its study-summary component computes results from these reviewed records under the specified monitoring rules. Historical token rules, remediation scope, and transfer purposes are verified inputs; outputs retain their sources. Appendix~\ref{app:D.1} details the two components and their input requirements.

Validation separates consistency with the research implementation from historical-state recovery. All 170 candidate classifications match the results of the original program. This comparison checks whether packaging the analysis as a toolkit preserved the original decisions. Separately, reconstructed allowances and balances are compared with saved historical on-chain query values. Validation inputs are fixed before execution, with expected results stored separately.

Five historical scenario sets reproduce renewed extraction, permission resetting, path closure, same-block spending, and controls. The corresponding states, calls, and decisions under five monitoring rules are reproduced. In these windows, all 21 balance checks after the initial state and all 9 recoverable allowance checks matched historical query values. Appendix~\ref{app:D.2} and Table~\ref{tab:D1} retain operation counts, complex-call checks, and individual comparisons. The toolkit runs offline using the Python standard library once local inputs are prepared, without API keys. Selected case materials, synthetic examples, tests, and reproduction instructions are documented in Appendix~\ref{app:D.3}.

\section{Recovery Processes and Fund Activity After Approval Incidents}\label{sec:4}

We examine how existing permissions were handled and which permissions supported spending after funds returned. The six-case design set contains 403 fixed relationships and 187 replenishment episodes; a separate sample contains 170 post-incident log candidates. The fixed relationships track changes in permissions and funds, whereas the separately drawn event sample identifies the operations represented by post-incident approval logs. External validation comprises one Dexible relationship and 20 Multichain relationships, reported separately. Each relationship links an account, token, and approved application. Spending by the application under study means that this application actually used the account's approval to transfer tokens.

\subsection{Remediation Measures and Permission Cleanup Progress}\label{sec:4.1}

Entry-point control and account permission cleanup progressed at different rates. The former changes the affected application function; explicitly setting an allowance to zero removes the account's transfer permission. We therefore verified when controls took effect separately from when account cleanup was completed. Socket and LiFi controlled their affected entry points within minutes of disclosure; Maestro and Dolomite did so before disclosure. Appendix~\ref{app:A} gives exact control times and components. Unibot announced a service suspension \cite{ref04}, while Sushi issued interface remediation and revocation notices \cite{ref14,ref15}. On-chain closure remains unverified for Unibot's affected entry point, and Sushi's interface-remediation time remains undetermined. Announcements and observed controls are therefore recorded separately. Account-side cleanup is counted from the allowance-zeroing operations actually observed.

Dexible paused its entry point before the public alert \cite{ref16}, as its team later recounted \cite{ref17}. Code, state, and control events support continued closure through day 30. Multichain's materials distinguish pool-system repairs from legacy-contract approval revocation \cite{ref05,ref06,ref18}. Closure evidence remains unavailable for the legacy contract studied here; Appendix~\ref{app:E} documents both external cases.

Cleanup objectives also differ. Applications patched at their original addresses may retain approvals for other functions; deprecated contracts require checking each relevant approval for revocation. We measure the first explicit zero setting against the fixed group with initially positive allowances, separately from consumption to zero or current zero states.

\begin{figure*}[tp]
\centering
\includegraphics[width=\textwidth,height=0.82\textheight,keepaspectratio]{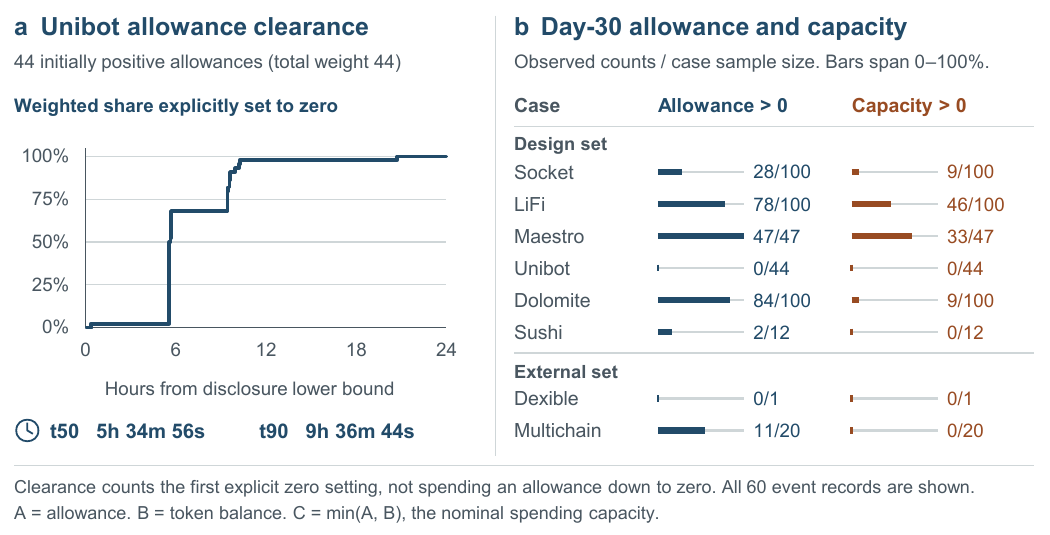}
\renewcommand{\thefigure}{2}
\caption{Allowance clearance and day-30 states. (a) Weighted share of Unibot's 44 initially positive allowances explicitly set to zero at least once. (b) Bars show within-case proportions; labels give counts and denominators. External cases are shown separately.}\label{fig:2}
\end{figure*}

All 44 initially positive Unibot relationships eventually underwent explicit zero settings. Weighted cleanup reached half at 5 hours 34 minutes 56 seconds after disclosure and nine-tenths at 9 hours 36 minutes 44 seconds. The last setting occurred at 20 hours 43 minutes 32 seconds. Figure~\ref{fig:2}a shows the complete trajectory: preexisting permissions remained during hours of account-by-account cleanup. Appendix~\ref{app:A} gives the first-setting time and first-to-last interval. Figure~\ref{fig:A3} in Appendix~\ref{app:A.4} preserves the complete cross-case cleanup comparison, including cases that did not reach the specified proportions.

The evidence places Multichain's disclosure time within a 74-hour interval. Its fixed initial group contains 19 relationships with positive allowances before the lower bound; 17 still have positive allowances at the upper-bound start. Subsequent explicit zero settings cleared 6 of these allowances, yielding approximately 35.2\% weighted cleanup. Including 2 settings within the disclosure interval raises this to approximately 47.0\%, using the same initial denominator and weights. Events within the disclosure interval remain in the sample rather than being dropped when the observation start changes. Neither calculation reached the half or nine-tenths cleanup threshold within 30 days. By day 30, 11 relationships retained allowances but had zero balances; nominal spending capacity was zero for all 20 relationships. These accounts temporarily held no funds, while the legacy application retained permission to transfer them. Appendix~\ref{app:E} gives the time alignment. Dexible's initially positive group is empty, so cleanup quantiles do not apply. Table~\ref{tab:2} summarizes the case-specific outcomes.

\subsection{End-of-Window Permission States and Approval Operations}\label{sec:4.2}

Figure~\ref{fig:2}b separates remaining allowances from current spending capacity. Of the 403 design-set relationships, 97 retained both allowances and balances on day 30, all in four cases with verified affected-entry-point closure. Closure covered specific functions, while account permissions could remain available to other functions.

\RQTakeaway{Closeout requires approval-level evidence}{%
\textbf{Finding.} Entry-point control and account permission cleanup progressed at different rates.
Of 403 relationships, 97 retained allowances and balances at day 30 despite verified closure of their affected entry points.
\par\smallskip
\textbf{Implication.} Responders should distinguish cleared allowances, approvals covered by verified entry-point closure, and relationships still lacking exit grounds.
With evidence available at the decision time, monitors can end emergency tracking of the original issue for eligible relationships.
Further checks and revocation reminders can then focus on those still requiring action.
}

\begin{table*}[!tp]
\centering\footnotesize
\renewcommand{\thetable}{2}
\caption{Allowance clearance and subsequent fund activity.}\label{tab:2}
\setlength{\tabcolsep}{4pt}\renewcommand{\arraystretch}{1.18}
\vbox{\offinterlineskip
\ProofRuleLine{ProofInk}{0.8pt}
\ProofBand{0pt}{\textwidth}{ProofBlueWash}{2.5pt}{1.6pt}{\begin{tabular}{>{\raggedright\arraybackslash}p{\dimexpr 0.160\textwidth-2\tabcolsep\relax}>{\raggedleft\arraybackslash}p{\dimexpr 0.133\textwidth-2\tabcolsep\relax}>{\raggedleft\arraybackslash}p{\dimexpr 0.165\textwidth-2\tabcolsep\relax}>{\raggedleft\arraybackslash}p{\dimexpr 0.168\textwidth-2\tabcolsep\relax}>{\raggedleft\arraybackslash}p{\dimexpr 0.168\textwidth-2\tabcolsep\relax}>{\raggedleft\arraybackslash}p{\dimexpr 0.206\textwidth-2\tabcolsep\relax}}
\ProofHead{Case} & \ProofHead{\shortstack[r]{Relationships\\\(n\)}} & \ProofHead{\shortstack[r]{Initially positive\\allowance}} & \ProofHead{\shortstack[r]{First explicit\\zero setting}} & \ProofHead{\shortstack[r]{Day-30 positive\\allowance}} & \ProofHead{\shortstack[r]{Replenishment\\episodes}} \\
\end{tabular}}
\hbox to\textwidth{\rlap{\color{ProofRule}\vrule width\textwidth height0.35pt depth0pt}\hskip\dimexpr.293\textwidth+\tabcolsep\relax{\color{ProofInk}\vrule width\dimexpr.501\textwidth-2\tabcolsep\relax height0.35pt depth0pt}\hss}
\ProofBand{0pt}{\textwidth}{white}{1.5pt}{1.5pt}{\begin{tabular}{>{\raggedright\arraybackslash}p{\dimexpr 0.160\textwidth-2\tabcolsep\relax}>{\raggedleft\arraybackslash}p{\dimexpr 0.133\textwidth-2\tabcolsep\relax}>{\raggedleft\arraybackslash}p{\dimexpr 0.165\textwidth-2\tabcolsep\relax}>{\raggedleft\arraybackslash}p{\dimexpr 0.168\textwidth-2\tabcolsep\relax}>{\raggedleft\arraybackslash}p{\dimexpr 0.168\textwidth-2\tabcolsep\relax}>{\raggedleft\arraybackslash}p{\dimexpr 0.206\textwidth-2\tabcolsep\relax}}
\multicolumn{6}{@{}l}{\textcolor{ProofInk}{\textbf{Design set}}}\\[3pt]
Socket & 100 & 37 & 9 & 28 & 41 \\
LiFi & 100 & 81 & 2 & 78 & 39 \\
Maestro & 47 & 45 & 0 & 47 & 13 \\
Unibot & 44 & 44 & 44 & 0 & 76 \\
Dolomite & 100 & 85 & 1 & 84 & 6 \\
Sushi RP2 & 12 & 5 & 3 & 2 & 12 \\
\end{tabular}}
\ProofRuleLine{ProofRule}{0.35pt}
\ProofBand{0pt}{\textwidth}{white}{1.5pt}{1.5pt}{\begin{tabular}{>{\raggedright\arraybackslash}p{\dimexpr 0.160\textwidth-2\tabcolsep\relax}>{\raggedleft\arraybackslash}p{\dimexpr 0.133\textwidth-2\tabcolsep\relax}>{\raggedleft\arraybackslash}p{\dimexpr 0.165\textwidth-2\tabcolsep\relax}>{\raggedleft\arraybackslash}p{\dimexpr 0.168\textwidth-2\tabcolsep\relax}>{\raggedleft\arraybackslash}p{\dimexpr 0.168\textwidth-2\tabcolsep\relax}>{\raggedleft\arraybackslash}p{\dimexpr 0.206\textwidth-2\tabcolsep\relax}}
\multicolumn{6}{@{}l}{\textcolor{ProofInk}{\textbf{External validation}}}\\[3pt]
Dexible & 1 & 0 & 0 & 0 & 0 \\
Multichain & 20 & 19 & 6 + 2\(\dagger\) & 11 & 3 \\
\end{tabular}}
\ProofRuleLine{ProofInk}{0.65pt}
}
\par\smallskip
\begin{minipage}{\textwidth}\footnotesize\raggedright
First zero settings use the initially positive group and exclude consumption. Multichain's two within-interval settings (\(\dagger\)) remain in the initial denominator. Episodes may repeat per relationship.
\end{minipage}
\end{table*}

Among these 97 relationships, 96 had pre-disclosure allowances with no subsequent positive setting or explicit zero setting through the end of observation. They represent continuation of preexisting permissions. The remaining Maestro relationship began with zero allowance and received a later positive setting. Thus, identical month-end states can represent continued or newly established permissions. Appendix~\ref{app:C.2} reports allowance magnitudes and residual USDC capacity.

The separate RQ2 event sample yielded 152 positive allowance settings and 18 consumption records among 170 candidates; none remained unclassified. For Unibot, only 2 of 20 candidates were settings. The other 18 spanned 7 transactions: the UNIBOT contract emitted an Approval log containing the remaining allowance after spending. These records primarily document use of existing permissions, rather than renewed approvals. All sampled Socket, LiFi, and Maestro candidates were settings. Table~\ref{tab:3} retains their bounded event populations and estimates, separately from the fixed relationships.

\begin{table*}[!tp]
\centering\footnotesize
\renewcommand{\thetable}{3}
\caption{Positive-allowance logs: settings versus consumption.}\label{tab:3}
\setlength{\tabcolsep}{4pt}\renewcommand{\arraystretch}{1.18}
\vbox{\offinterlineskip
\ProofRuleLine{ProofInk}{0.8pt}
\ProofBand{0pt}{\textwidth}{ProofBlueWash}{2.5pt}{1.6pt}{\begin{tabular}{>{\raggedright\arraybackslash}p{\dimexpr 0.158\textwidth-2\tabcolsep\relax}>{\raggedleft\arraybackslash}p{\dimexpr 0.102\textwidth-2\tabcolsep\relax}>{\raggedleft\arraybackslash}p{\dimexpr 0.083\textwidth-2\tabcolsep\relax}>{\raggedleft\arraybackslash}p{\dimexpr 0.132\textwidth-2\tabcolsep\relax}>{\raggedleft\arraybackslash}p{\dimexpr 0.150\textwidth-2\tabcolsep\relax}>{\raggedleft\arraybackslash}p{\dimexpr 0.105\textwidth-2\tabcolsep\relax}>{\raggedleft\arraybackslash}p{\dimexpr 0.270\textwidth-2\tabcolsep\relax}}
\ProofHead{Case} & \ProofHead{\(N\)} & \ProofHead{\(n\)} & \ProofHead{\shortstack[r]{Positive\\settings}} & \ProofHead{Consumption} & \ProofHead{Unknown} & \ProofHead{\shortstack[r]{Setting estimate\\(95\% interval)}} \\
\end{tabular}}
\ProofRuleLine{ProofRule}{0.35pt}
\ProofBand{0pt}{\textwidth}{white}{1.5pt}{0pt}{\begin{tabular}{>{\raggedright\arraybackslash}p{\dimexpr 0.158\textwidth-2\tabcolsep\relax}>{\raggedleft\arraybackslash}p{\dimexpr 0.102\textwidth-2\tabcolsep\relax}>{\raggedleft\arraybackslash}p{\dimexpr 0.083\textwidth-2\tabcolsep\relax}>{\raggedleft\arraybackslash}p{\dimexpr 0.132\textwidth-2\tabcolsep\relax}>{\raggedleft\arraybackslash}p{\dimexpr 0.150\textwidth-2\tabcolsep\relax}>{\raggedleft\arraybackslash}p{\dimexpr 0.105\textwidth-2\tabcolsep\relax}>{\raggedleft\arraybackslash}p{\dimexpr 0.270\textwidth-2\tabcolsep\relax}}
Socket & 1,754 & 50 & 50 & 0 & 0 & 1,754 (1,450--1,754) \\
LiFi & 8,323 & 50 & 50 & 0 & 0 & 8,323 (6,932--8,323) \\
Maestro & 59 & 50 & 50 & 0 & 0 & 59 (55--59) \\
\end{tabular}}
\ProofBand{0pt}{\textwidth}{ProofAmberWash}{0pt}{0pt}{\begin{tabular}{>{\raggedright\arraybackslash}p{\dimexpr 0.158\textwidth-2\tabcolsep\relax}>{\raggedleft\arraybackslash}p{\dimexpr 0.102\textwidth-2\tabcolsep\relax}>{\raggedleft\arraybackslash}p{\dimexpr 0.083\textwidth-2\tabcolsep\relax}>{\raggedleft\arraybackslash}p{\dimexpr 0.132\textwidth-2\tabcolsep\relax}>{\raggedleft\arraybackslash}p{\dimexpr 0.150\textwidth-2\tabcolsep\relax}>{\raggedleft\arraybackslash}p{\dimexpr 0.105\textwidth-2\tabcolsep\relax}>{\raggedleft\arraybackslash}p{\dimexpr 0.270\textwidth-2\tabcolsep\relax}}
Unibot & 20 & 20 & \ProofPair{2} & \ProofPair{18} & 0 & 2 (2--2) \\
\end{tabular}}
\ProofBand{0pt}{\textwidth}{white}{0pt}{1.5pt}{\begin{tabular}{>{\raggedright\arraybackslash}p{\dimexpr 0.158\textwidth-2\tabcolsep\relax}>{\raggedleft\arraybackslash}p{\dimexpr 0.102\textwidth-2\tabcolsep\relax}>{\raggedleft\arraybackslash}p{\dimexpr 0.083\textwidth-2\tabcolsep\relax}>{\raggedleft\arraybackslash}p{\dimexpr 0.132\textwidth-2\tabcolsep\relax}>{\raggedleft\arraybackslash}p{\dimexpr 0.150\textwidth-2\tabcolsep\relax}>{\raggedleft\arraybackslash}p{\dimexpr 0.105\textwidth-2\tabcolsep\relax}>{\raggedleft\arraybackslash}p{\dimexpr 0.270\textwidth-2\tabcolsep\relax}}
Dolomite & 0 & 0 & --- & --- & --- & 0 (empty frame) \\
Sushi RP2 & 0 & 0 & --- & --- & --- & 0 (empty frame) \\
\end{tabular}}
\ProofRuleLine{ProofInk}{0.65pt}
}
\par\smallskip
\begin{minipage}{\textwidth}\footnotesize\raggedright
N: candidate frame; n: audited sample. Estimates and 95\% intervals are case-specific. Dashes indicate inapplicable classifications in empty frames; external histories are excluded.
\end{minipage}
\end{table*}

Dolomite and Sushi had no candidates within their specified token, application, and window. Sushi's WETH can be spent through an approval without emitting a consumption-related Approval log. Actual spending therefore requires call-record checks even when the candidate frame is empty. The absence of positive-allowance logs and the absence of application spending are separate observations because the token determines which operations produce these records.

Accounts continued to set positive allowances for applications after remediation at their original addresses. All 50 sampled Socket settings and all 50 Maestro settings followed remediation; LiFi had 48 afterward and 2 beforehand. Unibot's 2 settings preceded its suspension announcement, the reference time for this case; subsequent candidates recorded consumption. Across the 21 external relationships, no positive overwrites occurred during the primary window. Multichain had only the explicit zero settings reported above, and Dexible had no Approval logs. These relationship histories retain a different denominator from the 170 event candidates.

\subsection{Permission Use After Funds Return}\label{sec:4.3}

\begin{figure*}[tp]
\centering
\includegraphics[width=\textwidth,height=0.82\textheight,keepaspectratio]{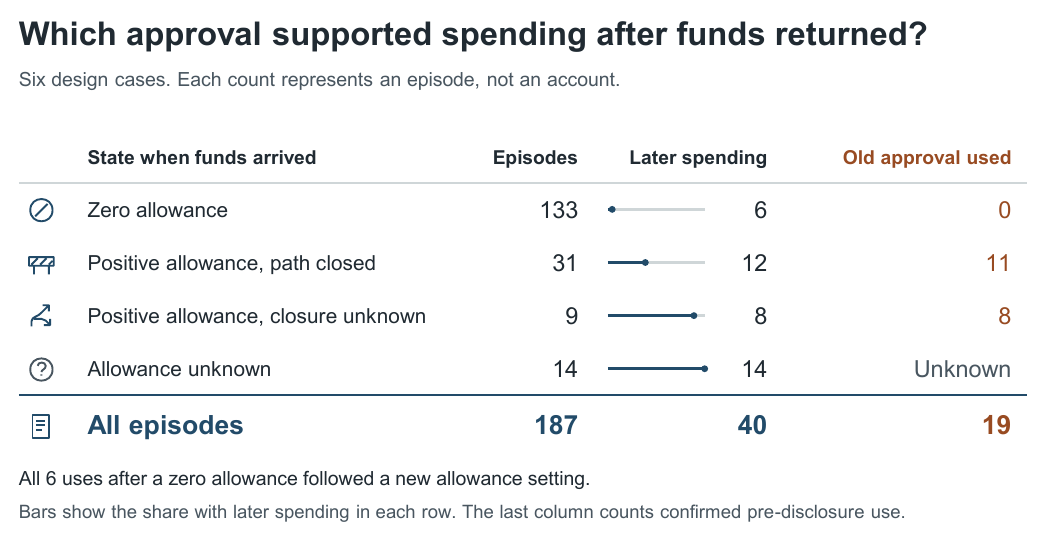}
\renewcommand{\thefigure}{3}
\caption{Permission sources across 187 replenishment episodes. Bars show the share of episodes in each row with later spending by the original spender. The final column counts confirmed uses of pre-disclosure approvals; unknown sources remain separate.}\label{fig:3}
\end{figure*}

After funds return, application spending may rely on a permission retained from before disclosure or one established through a later positive setting. Among the 403 fixed relationships, 85 experienced zero-to-positive balance transitions, producing 187 replenishment episodes. Of these, 40 involved subsequent spending by the application under study, including 19 confirmed uses of pre-disclosure allowances. We confirmed spending from actual calls, traced the supporting permission history, and checked the affected entry point's control state. Figure~\ref{fig:3} groups episodes by their conditions at inflow; one relationship can contribute several episodes.

All 6 spending episodes that began with zero allowances involved positive settings after inflow and before spending. These episodes, drawn from 133 zero-allowance episodes, involved 3 relationships. Permissions were reestablished before use.

Of the 31 episodes beginning with positive allowances and verified closure, 12 involved subsequent spending; 11 used pre-disclosure permissions. Patching a specific entry point left approvals available to other functions. Checking both calls and permission history separates continued use of a preexisting approval from continued use of the affected entry point. Another 9 episodes began with positive allowances but unknown closure status; 8 involved spending, all under pre-disclosure allowances. Together, these groups account for the 19 confirmed uses: 11 with closed paths and 8 with unknown closure. Permission source alone does not establish whether a transfer was account-initiated or third-party-directed, nor does it determine the transfer's purpose. Activity and purpose labels are assessed separately using transaction evidence in Section~\ref{sec:4.4} and Appendix~\ref{app:B}.

The 14 unknown-allowance episodes all came from one LiFi relationship. Actual calls establish spending, but the allowance at inflow could not be reconstructed precisely. Their spending observations remain in the results even though the permission state at the start of each episode is unresolved. These episodes retain unknown status; Appendix~\ref{app:B} explains the technical reason. Unsupported entry-point states and purposes likewise remain unknown.

Of the 187 episodes, 147 had no subsequent application spending. In 42, balances remained positive at the end of observation. These categories can overlap: the first concerns spending, whereas the second concerns whether the balance episode ended.

By day 30, nominal spending capacity was zero in 75 of the 85 relationships with replenishment. Across all 403 relationships, 49 involved application spending during the full window; 31 of these had zero day-30 capacity. These 49 are relationships with spending anywhere in the 30-day window, without requiring replenishment. The 40 spending episodes instead belong to the 187 replenishment episodes. Month-end states therefore cannot replace transaction histories when determining whether permissions were used during the month.

\subsection{Key Fund Episodes and External Protocol Validation}\label{sec:4.4}

\textbf{Unibot: extraction before permission cleanup.} One wallet repurchased UNIBOT twice and experienced repeated extraction 60 and 84 seconds later, respectively. Neither interval contained a positive allowance setting; both transfers used the preexisting permission. The contemporaneous PeckShield alert matches the address, amounts, and sequence \cite{ref10}. The allowance was explicitly set to zero afterward, connecting hours-long cleanup to fund consequences within minutes.

These records, U2 and U3, belong to Unibot's 76 replenishment episodes. Of 5 episodes beginning with positive allowances and unknown closure status, 4 involved pre-disclosure allowance use, including U2 and U3. The other 71 began with zero allowances and had no application spending during their respective episodes. Among the observed Unibot episodes, pre-disclosure allowance use occurred only in episodes that began with positive allowances. Appendix~\ref{app:B} preserves complete times, amounts, transaction references, and the other illustrative episodes.

\begin{figure*}[tp]
\centering
\includegraphics[width=\textwidth,height=0.82\textheight,keepaspectratio]{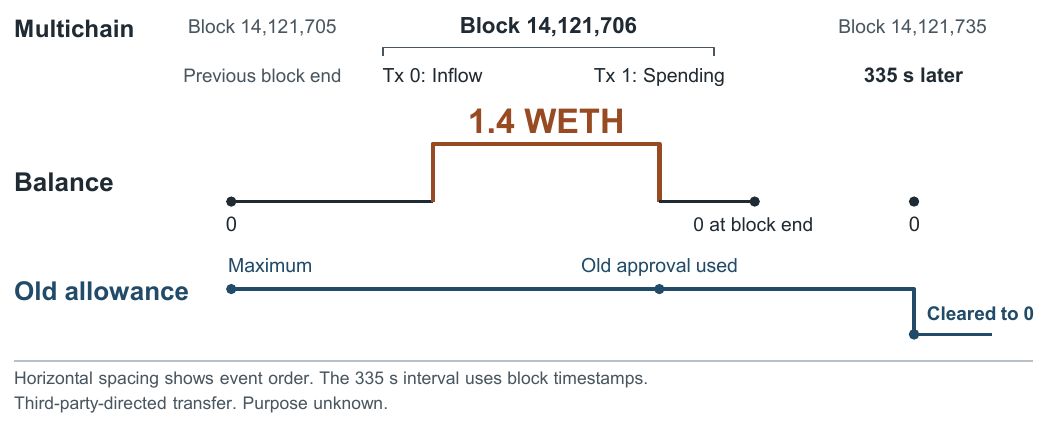}
\renewcommand{\thefigure}{4}
\caption{Balance and allowance for one Multichain relationship. Adjacent transactions brought in and spent 1.4 WETH, leaving zero balances at both block boundaries. The pre-disclosure allowance remained at its maximum until it was set to zero 335 seconds later. The transfer was third-party-directed; its purpose remains unknown.}\label{fig:4}
\end{figure*}

\textbf{Multichain: spending between zero-balance observations.} In one external relationship, 1.4 WETH entered and left through two adjacent transactions in the same block. The legacy contract spent it under a pre-disclosure allowance. Both boundary balances were zero, concealing the intervening inflow and use from endpoint observations. Figure~\ref{fig:4} reconstructs the transaction order.

The spending occurred more than 13 days after the disclosure interval ended, with no intervening positive allowance setting. Execution records identify the legacy contract and the same entry point as the reference incident. WETH does not decrement this maximum allowance on use, so permission remained after the funds left. The transfer retains its third-party-directed label, with purpose unknown. Appendix~\ref{app:E} provides the state, call, and timing evidence.

Multichain contributed 3 episodes from 2 relationships. One was the pre-disclosure use described above; the other relationship began with zero allowance and contributed 2 inflows without application spending. Their different permission histories distinguish their outcomes.

\textbf{Dexible: outflows without application spending.} Its sole TRU relationship had two balance outflows but no spending by the application under study during the formal window. Allowance was already zero before disclosure; the reference incident transaction also records allowance consumption to zero. The entry point had verified closure, and no new Approval logs or replenishment episodes occurred. Post-incident outflows therefore require two checks: whether the application under study spent the funds, and which approval supported that spending.

\RQTakeaway[H]{Later spending must be traced\newline to its supporting approval}{%
\textbf{Finding.} Of 40 spending episodes after funds returned, 19 were confirmed to use pre-disclosure approvals.
All six spending episodes that began with zero allowance involved a later setting before spending.
\par\smallskip
\textbf{Implication.} Old permissions without verified exit grounds should remain under monitoring even while accounts temporarily hold no funds.
After emergency monitoring ends, retain approval and control-change records and reassess when new approvals or relevant control changes appear.
Investigators should first establish whether the application spent the funds, then identify the supporting approval to distinguish retained permissions from later settings.
}

\section{Criteria for Incident Closeout and Practical Recommendations}\label{sec:5}

After a project team implements remediation, monitors must determine how that response applies to the approvals they are tracking. This leaves a relationship-level decision: which approvals still require monitoring for the original issue, and which have grounds for exit? Current funds, remaining permissions, and entry-point controls answer different questions. We compare their decision criteria on identical historical records.

\subsection{Monitoring Rules and Comparison Conditions}\label{sec:5.1}

Under each information condition, all rules read the same allowance, balance, and control evidence, but apply different grounds for retaining a specific approval. The balance rule asks whether the account currently holds funds and exits when balance is confirmed zero. The spendable-amount rule (capacity rule) asks how much the application can currently transfer, jointly considering balance and allowance. It exits when either is confirmed zero. The approval rule (R0) asks whether transfer permission remains and exits when allowance is confirmed zero; temporary absence of funds does not affect retention. The approval-and-remediation rule (R1) additionally exits when admissible evidence confirms closure of the relationship's affected entry point.

A full-evidence review (F) checks whether the materials support further retention or exit reasons beyond R1. We use its decisions to check whether R1's simpler conditions capture every retention or exit action supported by the available evidence. Unknown states follow each rule's predefined conditions in Appendix~\ref{app:C.5}.

Four checks, at disclosure and days 1, 7, and 30, produce 1,612 relationship--checkpoint decisions for 403 relationships. A relationship retained throughout contributes four decisions, not four accounts. The 21 external relationships contribute 84 decisions, reported separately. Episode comparisons use verified end-of-block states; pre-spending decisions use the latest observation strictly before the spending block.

Retrospective interpretation (H) uses control interpretations verified in the retrospective analysis. Contemporaneously available evidence (O) admits only evidence confirmed to have been available at the decision time. The first asks what grounds for a decision can be established in hindsight. The second asks what evidence was verifiably available when a decision was needed. Additional scenarios assume hypothetical evidence-formation delays of 0, 1, and 24 hours. Table~\ref{tab:4} reports all scenarios; Appendix~\ref{app:C.5} specifies evidence-use and update rules.

\begin{table*}[!t]
\centering\footnotesize
\renewcommand{\thetable}{4}
\caption{Monitoring decisions under matched information conditions.}\label{tab:4}
\setlength{\tabcolsep}{4pt}\renewcommand{\arraystretch}{1.22}
\vbox{\offinterlineskip
\ProofRuleLine{ProofInk}{0.8pt}
\ProofBand{0pt}{\textwidth}{ProofBlueWash}{2.5pt}{1.6pt}{\begin{tabular}{>{\raggedright\arraybackslash}p{\dimexpr 0.324\textwidth-2\tabcolsep\relax}>{\raggedleft\arraybackslash}p{\dimexpr 0.095\textwidth-2\tabcolsep\relax}>{\raggedleft\arraybackslash}p{\dimexpr 0.107\textwidth-2\tabcolsep\relax}>{\raggedleft\arraybackslash}p{\dimexpr 0.134\textwidth-2\tabcolsep\relax}>{\raggedleft\arraybackslash}p{\dimexpr 0.174\textwidth-2\tabcolsep\relax}>{\raggedleft\arraybackslash}p{\dimexpr 0.166\textwidth-2\tabcolsep\relax}}
\ProofHead{Information condition} & \ProofHead{Balance} & \ProofHead{Capacity} & \ProofHead{Approval} & \ProofHead{Approval +} & \ProofHead{Full review} \\
 &  &  & \ProofHead{(R0)} & \ProofHead{closure (R1)} & \ProofHead{(F)} \\
\end{tabular}}
\ProofRuleLine{ProofRule}{0.35pt}
\ProofBand{0pt}{\textwidth}{white}{1.5pt}{0pt}{\begin{tabular}{>{\raggedright\arraybackslash}p{\dimexpr 0.324\textwidth-2\tabcolsep\relax}>{\raggedleft\arraybackslash}p{\dimexpr 0.095\textwidth-2\tabcolsep\relax}>{\raggedleft\arraybackslash}p{\dimexpr 0.107\textwidth-2\tabcolsep\relax}>{\raggedleft\arraybackslash}p{\dimexpr 0.134\textwidth-2\tabcolsep\relax}>{\raggedleft\arraybackslash}p{\dimexpr 0.174\textwidth-2\tabcolsep\relax}>{\raggedleft\arraybackslash}p{\dimexpr 0.166\textwidth-2\tabcolsep\relax}}
\multicolumn{6}{@{}l}{\textcolor{ProofInk}{\textbf{Design set}}\quad 1,612 relationship--checkpoints}\\[3pt]
\end{tabular}}
\ProofBand{0pt}{\textwidth}{ProofBlueWash}{0pt}{1.5pt}{\begin{tabular}{>{\raggedright\arraybackslash}p{\dimexpr 0.324\textwidth-2\tabcolsep\relax}>{\raggedleft\arraybackslash}p{\dimexpr 0.095\textwidth-2\tabcolsep\relax}>{\raggedleft\arraybackslash}p{\dimexpr 0.107\textwidth-2\tabcolsep\relax}>{\raggedleft\arraybackslash}p{\dimexpr 0.134\textwidth-2\tabcolsep\relax}>{\raggedleft\arraybackslash}p{\dimexpr 0.174\textwidth-2\tabcolsep\relax}>{\raggedleft\arraybackslash}p{\dimexpr 0.166\textwidth-2\tabcolsep\relax}}
H: retrospective evidence & 592 & 412 & \ProofKeep{1,023} & \ProofKeep{176} & \ProofKeep{176} \\
O: contemporaneously available evidence & 592 & 412 & \ProofKeep{1,023} & \ProofKeep{1,023} & \ProofKeep{1,023} \\
\end{tabular}}
\ProofRuleLine{ProofRule}{0.25pt}
\ProofBand{0pt}{\textwidth}{white}{1.5pt}{1.4pt}{\begin{tabular}{>{\raggedright\arraybackslash}p{\dimexpr 0.324\textwidth-2\tabcolsep\relax}>{\raggedleft\arraybackslash}p{\dimexpr 0.095\textwidth-2\tabcolsep\relax}>{\raggedleft\arraybackslash}p{\dimexpr 0.107\textwidth-2\tabcolsep\relax}>{\raggedleft\arraybackslash}p{\dimexpr 0.134\textwidth-2\tabcolsep\relax}>{\raggedleft\arraybackslash}p{\dimexpr 0.174\textwidth-2\tabcolsep\relax}>{\raggedleft\arraybackslash}p{\dimexpr 0.166\textwidth-2\tabcolsep\relax}}
Hypothetical 0-hour delay & 592 & 412 & 1,023 & 176 & 176 \\
Hypothetical 1-hour delay & 592 & 412 & 1,023 & 306 & 306 \\
Hypothetical 24-hour delay & 592 & 412 & 1,023 & 416 & 416 \\
\end{tabular}}
\ProofRuleLine{ProofRule}{0.35pt}
\ProofBand{0pt}{\textwidth}{white}{1.6pt}{1.5pt}{\begin{tabular}{>{\raggedright\arraybackslash}p{\dimexpr 0.324\textwidth-2\tabcolsep\relax}>{\raggedleft\arraybackslash}p{\dimexpr 0.095\textwidth-2\tabcolsep\relax}>{\raggedleft\arraybackslash}p{\dimexpr 0.107\textwidth-2\tabcolsep\relax}>{\raggedleft\arraybackslash}p{\dimexpr 0.134\textwidth-2\tabcolsep\relax}>{\raggedleft\arraybackslash}p{\dimexpr 0.174\textwidth-2\tabcolsep\relax}>{\raggedleft\arraybackslash}p{\dimexpr 0.166\textwidth-2\tabcolsep\relax}}
\multicolumn{6}{@{}l}{\textcolor{ProofInk}{\textbf{External validation}}\quad 84 relationship--checkpoints}\\[3pt]
Identical decisions in all five scenarios & 10 & 2 & 59 & 59 & 59 \\
\end{tabular}}
\ProofRuleLine{ProofInk}{0.65pt}
}
\par\smallskip
\begin{minipage}{\textwidth}\footnotesize\raggedright
Entries count retained relationship--checkpoints. H uses retrospective evidence; O uses evidence confirmed available at the time. Evidence-availability delays are hypothetical.
\end{minipage}
\end{table*}

\subsection{Differences Produced by the Decision Criteria}\label{sec:5.2}

\textbf{An empty account can retain relevant permission.} Before Multichain's same-block spending, the balance and capacity rules had no positive-balance observation and did not retain the relationship. R0, R1, and F retained it because permission remained. For Unibot's cross-block U2/U3 episodes, all four rules and F retained the relationship before spending. R0/R1/F did so continuously; the fund-based rules included or reincluded it when funds arrived. Under the study's block-level updates, the difference turns on whether a positive balance is observable between inflow and spending. At Multichain's day-30 checkpoint, capacity retained none of 20 relationships, whereas R0 retained 11: current spendable amounts and unremoved permissions support different decisions.

\textbf{Applicable closure evidence provides exit grounds.} The 97 day-30 relationships with both allowances and balances had verified closure of their affected entry points. Mapping control scope to each relationship allows exit from monitoring the original issue despite retained permissions. Under H, adding this evidence changes R0's 1,023 retained items to R1's 176. The 847 changes are retain-to-exit decisions measured in relationship--checkpoints across four checks of 403 relationships, not counts of unique accounts. Verifiable control evidence thus identifies relationships that can leave the emergency list, focusing attention on those still lacking exit grounds. Across all five design-set scenarios, the balance and capacity rules retain 592 and 412 decisions, respectively. Their fund-based criteria do not use the closure interpretations whose availability changes between scenarios.

R1 and F each retain 176 items because the full review supplies no additional retention or exit reasons. In this dataset, two checks capture the supported actions: whether the allowance is zero or applicable closure evidence is available. Either condition provides exit grounds; otherwise, the relationship remains under monitoring. Reliable permission records and control verification remain necessary inputs. In the external group, R0/R1/F each retain 59 of 84 decisions under all five scenarios, for different reasons. Dexible's allowance was already zero, so closure adds no exit opportunity. Multichain lacks applicable closure evidence, so no additional path-based exit is justified.

\textbf{Evidence availability determines when exit is justified.} Under O, R0/R1/F all retain 1,023 decisions. Retrospective evidence identifies when the original incident path was closed. A contemporaneous exit decision requires closure evidence that was available at that time. Hypothetical delays of 0, 1, and 24 hours yield 176, 306, and 416 retained decisions for both R1 and F. Later availability leaves more decisions in the retained state. Table~\ref{tab:4} reports these results, and Figure~\ref{fig:A5} in Appendix~\ref{app:C.5} preserves the complete graphical comparison. Remediation records must distinguish the measure's effective time from the time its evidence became verifiable, while identifying the covered entry point and relationships.

\subsection{Specific Measures for Incident Response and Approval Monitoring}\label{sec:5.3}

Emergency monitoring lists should be managed separately from event records retained for reassessment. The list identifies relationships needing continued attention; the event records preserve the materials needed to reconsider decisions after permission or control changes. Wallet-side work addresses allowance recommendations and signing-time warnings \cite{ref19}; the measures here concern post-disclosure response and investigation.

\textbf{Remediation teams should specify coverage.} Reports should identify patched or closed functions, versions, covered approval relationships, effective times, and accounts still requiring revocation. They should record when control changes and interpretations became verifiable. Multichain's pool-system repair and legacy-contract revocation requirements concern different targets, whereas Dexible has verifiable entry-point closure evidence. Separating these targets lets monitors determine whether a reported measure actually covers the approval being assessed.

\textbf{Monitors should record reasons for retention and exit.} Monitoring of the legacy path ends when allowance is confirmed zero or applicable closure evidence available at the time supports exit. Positive or unknown allowances without applicable closure evidence remain retained. Each decision should preserve the relationship, observation time, and supporting evidence so its grounds can be checked.

\textbf{Permission or control changes should trigger reassessment.} The six initially zero-allowance episodes with later spending all had subsequent positive settings. Monitors should distinguish settings from consumption and continue recording relevant control changes. A confirmed positive setting calls for checking current permission and path states before deciding whether to include the relationship. Changes to a shared entry point require reassessing all affected relationships together. Initial inclusion and reinclusion after exit should be recorded separately, with updated decisions and reasons.

\textbf{Investigators should reconstruct transaction order.} Records should link inflows, actual transfers on behalf of accounts, allowance changes, and their sequence, including activity between zero-balance observations. Each transfer should be linked to its supporting permission and transaction-specific purpose evidence. Investigators can then determine whether it used a pre-disclosure approval and whether it passed through the affected entry point.

The Authorization Recovery Toolkit identifies operations, reconstructs episodes from saved transaction records, and recomputes monitoring decisions from reviewed materials, providing reusable steps for these checks.

\subsection{Robustness and Scope of the Conclusions}\label{sec:5.4}

Moving Multichain's start to the disclosure interval's beginning changes early states and episode counts, but retains day-30 allowances and the same-block pre-disclosure use. Neither anchor reaches weighted cleanup \(t_{50}\) or \(t_{90}\); Table~\ref{tab:E2} details the counts and coverage.

Prior exposure means relationships previously examined or used for preliminary technical reads. Removing 25 such Unibot relationships reduces relationships with third-party spending after replenishment from 3 to 0, concentrating that observation in the prior-exposure group. Both external protocols had prior literature prescreening. Dexible's sole relationship overlaps a technical-read account; none of Multichain's 20 relationships does. Sections~\ref{app:C.3} and~\ref{app:C.4} of the supplementary material report design-set sensitivity to time anchors, residual-value thresholds, and protocol composition, including weighted results.

The Unibot comparison (4/5 versus 0/71) uses episodes from a single case. A relationship can contribute multiple episodes. Multichain's same-block use comes from one newly sampled relationship. Findings apply to selected protocols, tokens, and bounded frames; historical decision counts are not converted to labor hours, costs, or prevented losses.

We distinguish account-initiated activity from third-party-directed activity and assess transfer purposes separately. Specific business, attack, or rescue labels require transaction-specific evidence; unresolved purposes remain unknown. Analyses use historical records without broadcasting transactions, replaying attacks, or inferring natural-person identities. Appendix materials preserve UNKNOWN bounds and transaction-source indexes.

\section{Related Work}\label{sec:6}

Approval-incident recovery requires connecting project repairs to verifiable exit grounds for individual permissions. Four research areas provide the basis for this relationship-level analysis.

\subsection{Token Approvals, Permission Exposure, and User Interaction}\label{sec:6.1}

Wang et al. quantify nominal spending capacity from allowances and balances, examining unlimited approvals, future-inflow risks, and approval--execution sequences \cite{ref03}. OSPREY identifies controllable transferFrom vulnerabilities in approved applications and evaluates historical approvals and exploitable instances \cite{ref02}. Together, these studies establish how persistent permissions expose funds and how application flaws enable their abuse.

TxPhishScope visits suspicious websites and triggers and simulates transactions to identify transaction-based phishing \cite{ref20}. Chen et al. examine payloads that abuse legitimate contracts or invoke phishing contracts \cite{ref08}. Adamczyk studies approval-based investment fraud and wallet presentation and verification of approval information \cite{ref21}. Guan et al. compare allowance-limit recommendations, spender reminders, and delayed confirmation through user experiments \cite{ref19}. These studies connect dangerous permissions to transaction semantics and signing decisions. Our analysis extends approval-exposure assessment to recovery decisions by matching permission histories to remediation scope and subsequent fund use.

\subsection{Actual Execution, Token Semantics, and Fund Forensics}\label{sec:6.2}

Securify checks contract security properties \cite{ref22}, while teEther generates executable exploit witnesses \cite{ref23}. Perez and Livshits investigate actual historical exploitation of flagged contracts \cite{ref24}, and Chaliasos et al. evaluate security tools on real attack samples \cite{ref25}. These studies distinguish property checks, exploitability, and observed attacks as separate evidence levels.

TXSPECTOR extracts control and data dependencies through transaction replay for forensic rules \cite{ref26}. EthScope organizes fine-grained historical states for replay and attack investigation \cite{ref27}. Horus combines execution analysis with fund-flow tracing, including exchanges, returns, and recovery accounts \cite{ref28}. TokenScope checks consistency among balance storage, interfaces, and transfer events \cite{ref13}. SODA separates runtime information collection from detection applications for extensible online analysis \cite{ref29}. These capabilities establish execution facts and token semantics needed to reconstruct transfers. We connect those facts to complete permission histories and entry-point controls, distinguishing continued calls to one contract from use of the original incident path.

\subsection{Upgrades, Migration, Patching, and Scope of Control}\label{sec:6.3}

Salehi et al. compare upgrade mechanisms and administrative authority, including how migration and proxies handle states and entry points \cite{ref30}. Proxy Hunting examines proxy-upgrade design, implementation, and policy, including delayed emergency patching \cite{ref31}. EVMPatch repairs selected defects using bytecode rewriting and proxies \cite{ref32}. SmartUpdater studies logic and state maintenance in stateful contracts \cite{ref33}, while Sereum uses runtime checks to protect existing contracts against reentrancy \cite{ref34}. These methods change different programs, states, or execution conditions, making control scope central to remediation analysis.

Li et al. document continued use of old addresses after migration from explicitly deprecated contract versions \cite{ref35}. Qasse et al. relate upgrade histories, code changes, and contract activity \cite{ref36}. These findings show that version changes and user activity follow different trajectories. We separately verify project-level repair and account-level cleanup, interpreting subsequent spending through its actual entry point and permission source.

\subsection{Incident Response, Victim Trajectories, and Recovery Processes}\label{sec:6.4}

DeFi systematization organizes application components and technical and economic risks \cite{ref37}. Carpentier-Desjardins et al. categorize crime incidents by project role and technical layer \cite{ref38}. Zhou et al. measure emergency-pause timing \cite{ref07}. They distinguish the rescue window, from attack-contract deployment to first harm, from the incident window between first and last harm. Blockchain Imitation Game studies transaction imitation for preemptive rescue \cite{ref39}. Parhizkari et al. identify potentially affected addresses between deployment and attack execution \cite{ref40}.

Chen et al. classify phishing victims' responses as approval revocation, asset movement, or no action \cite{ref08}. Seo and Ahn characterize replenishment, repeated extraction, and victim trajectories, estimating delay-related loss bounds under idealized instantaneous blocking \cite{ref09}. Liu et al. use interviews and questionnaires to examine users' risk perceptions, responses, and post-incident behavior \cite{ref41}. We connect project repairs to approval-specific closeout by recording the controlled components, scope, and effective time of each control. We also record when the supporting evidence became available, retaining unknown times where necessary. These records support decisions to continue monitoring, end monitoring, or reassess a relationship.

\section{Conclusion}\label{sec:7}

Recovery requires verifiable closeout grounds for each preexisting approval, connecting project remediation to permission changes and fund activity. Project remediation and account permission cleanup progressed on different timelines. This pattern was observed across 403 fixed relationships in six cases over 30 days, with external verification in Dexible and Multichain. Repeated extraction during cleanup and same-block spending between zero-balance observations show why monitoring can remain necessary even when observed balances are zero. Conversely, scoped entry-point closure can support exit despite retained approvals.

Post-incident spending used both pre-disclosure permissions and permissions established through subsequent settings. Positive settings and allowance consumption represent different operations, and the same application can continue transferring funds under approvals established at different times. These histories connect subsequent fund use to its supporting permissions and the controls in effect.

Emergency monitoring of the original incident path ends when the current allowance is verified zero or applicable closure evidence is available at the decision time. New permission settings or control changes require the relationship to be reassessed. The historical comparison shows how evidence availability changes closeout decisions. Under the same information conditions, the simple conditions produced the same decisions as the full-evidence review. Project reports should specify controlled components, scope, effective time, and evidence availability. Monitors should preserve decision grounds; forensic and research teams can use the Authorization Recovery Toolkit to reconstruct activity and recompute decisions offline.

\onecolumn
\appendices
\emergencystretch=1em
\section{Case Scope, Incident Timing, and Remediation Evidence}\label{app:A}

\subsection{Public Disclosure and Actual Remediation}\label{app:A.1}

This study separately records when each incident became public, what measures the project took, and when those measures took effect. Public notices establish the observation start, while contract configurations and execution records establish the actual scope of control. All cases are on Ethereum mainnet, with chain ID 1.

Table~\ref{tab:A1} lists public disclosure times and control evidence for the six-case design set. Public records for Socket, LiFi, Maestro, and Unibot have second-level resolution, while those for Dolomite and Sushi have millisecond-level resolution. Each listed time is the lower bound of the corresponding record interval; the upper bound is the start of the next second or millisecond. The study uses this temporal resolution to locate observation boundaries. All times are in UTC.

\begin{table}[!htbp]
\centering\footnotesize
\renewcommand{\thetable}{A1}
\setlength{\tabcolsep}{4pt}
\renewcommand{\arraystretch}{1.16}
\caption{Public Disclosure Times, Remediation Actions, and Verification Evidence}\label{tab:A1}
\begin{tabular}{>{\raggedright\arraybackslash}p{\dimexpr 0.110000\textwidth-2\tabcolsep\relax}>{\raggedright\arraybackslash}p{\dimexpr 0.190000\textwidth-2\tabcolsep\relax}>{\raggedright\arraybackslash}p{\dimexpr 0.230000\textwidth-2\tabcolsep\relax}>{\raggedright\arraybackslash}p{\dimexpr 0.470000\textwidth-2\tabcolsep\relax}}
\toprule
\textbf{Case} & \textbf{Disclosure lower bound, UTC} & \textbf{Remediation action} & \textbf{Technical location and evidence status} \\
\midrule
Socket & 2024-01-16 19:19:49.000 & Disabled the affected route & Block 19021526; query results for route 406 agree with configuration-change events \\
LiFi & 2024-07-16 12:28:53.000 & Removed the affected function & Block 20319191; the implementation address query (facetAddress) for function selector \texttt{0x3d86ea\allowbreak{}87} agrees with configuration-change events \\
Maestro & 2023-10-25 00:07:47.000 & Upgraded the proxy contract's execution logic & Block 18423429; the implementation address storage location and call dispatch logic confirm that the old function \texttt{0x923912\allowbreak{}7f} was disabled \\
Unibot & 2023-10-31 05:18:39.000 & Publicly announced a service pause at 06:25:39 \cite{ref04} & Block 18468310 near the announcement; on-chain closure of the affected entry point is unknown \\
Dolomite & 2024-03-20 17:14:00.251 & Revoked call eligibility for the old entry point, then paused delegated execution & Eligibility revoked at block 19477294; delegate execution paused at block 19479508; the two controls are located separately \\
Sushi RP2 & 2023-04-09 03:36:53.830 & Interface remediation and revocation notices \cite{ref14,ref15} & The UTC time of interface remediation is unknown; the public postmortem documents the response context \\
\bottomrule
\end{tabular}
\end{table}

The control actions for Socket and LiFi occurred 5 minutes 58 seconds and 23 minutes 30 seconds after public disclosure, respectively. Maestro's upgrade preceded disclosure by 31 minutes 48 seconds; Dolomite's revocation of eligibility for the old entry point preceded disclosure by 6 minutes 37.251 seconds. Dolomite's two actions were 7 hours 26 minutes 12 seconds apart and separately controlled entry-point eligibility and delegated execution.

A project-affiliated Unibot account posted a retrospective statement at 07:01:04 on 2023-10-31 that revocations had been completed \cite{ref42}. The batch of accounts covered by this statement is recorded as unknown. We determine cleanup progress from the actual operation that set each sampled relationship's allowance to zero. Among the 44 relationships with initially positive allowances, zero-setting operations began 21 minutes 20 seconds after disclosure. The first and last operations were 20 hours 22 minutes 12 seconds apart, with the last occurring 20 hours 43 minutes 32 seconds after disclosure.

\subsection{Contract Identities and Amount Units}\label{app:A.2}

Each approval relationship involves a token and an application authorized to transfer it. Table~\ref{tab:A2} provides the complete contract addresses for the six cases to identify the study objects when reviewing transactions and historical states.

\begin{table}[!htbp]
\centering\footnotesize
\renewcommand{\thetable}{A2}
\setlength{\tabcolsep}{4pt}
\renewcommand{\arraystretch}{1.16}
\caption{Tokens and Approved Applications Used in the Study}\label{tab:A2}
\begin{tabular}{>{\raggedright\arraybackslash}p{\dimexpr 0.140000\textwidth-2\tabcolsep\relax}>{\raggedright\arraybackslash}p{\dimexpr 0.430000\textwidth-2\tabcolsep\relax}>{\raggedright\arraybackslash}p{\dimexpr 0.430000\textwidth-2\tabcolsep\relax}}
\toprule
\textbf{Case} & \textbf{Token contract address (token)} & \textbf{Approved application address (spender)} \\
\midrule
Socket & \texttt{0xa0b869\allowbreak{}91c6218b\allowbreak{}36c1d19d\allowbreak{}4a2e9eb0\allowbreak{}ce3606eb\allowbreak{}48} & \texttt{0x3a23f9\allowbreak{}43181408\allowbreak{}eac42411\allowbreak{}6af7b779\allowbreak{}0c94cb97\allowbreak{}a5} \\
LiFi & \texttt{0xa0b869\allowbreak{}91c6218b\allowbreak{}36c1d19d\allowbreak{}4a2e9eb0\allowbreak{}ce3606eb\allowbreak{}48} & \texttt{0x1231de\allowbreak{}b6f5749e\allowbreak{}f6ce6943\allowbreak{}a275a1d3\allowbreak{}e7486f4e\allowbreak{}ae} \\
Maestro & \texttt{0xaaee1a\allowbreak{}9723aadb\allowbreak{}7afa2810\allowbreak{}263653a3\allowbreak{}4ba2c21c\allowbreak{}7a} & \texttt{0x80a64c\allowbreak{}6d7f12c4\allowbreak{}7b7c66c5\allowbreak{}b4e20e72\allowbreak{}bc1fcd5d\allowbreak{}9e} \\
Unibot & \texttt{0xf819d9\allowbreak{}cb1c2a81\allowbreak{}9fd99178\allowbreak{}1a822de3\allowbreak{}ca8607c3\allowbreak{}c9} & \texttt{0x126c9f\allowbreak{}bab3a2fc\allowbreak{}a24edfd1\allowbreak{}7322e71a\allowbreak{}5e36e918\allowbreak{}65} \\
Dolomite & \texttt{0xa0b869\allowbreak{}91c6218b\allowbreak{}36c1d19d\allowbreak{}4a2e9eb0\allowbreak{}ce3606eb\allowbreak{}48} & \texttt{0xe2466d\allowbreak{}eb9536a6\allowbreak{}9bf8131e\allowbreak{}cd0c267e\allowbreak{}e41dd1cd\allowbreak{}a0} \\
Sushi RP2 & \texttt{0xc02aaa\allowbreak{}39b223fe\allowbreak{}8d0a0e5c\allowbreak{}4f27ead9\allowbreak{}083c756c\allowbreak{}c2} & \texttt{0x044b75\allowbreak{}f554b886\allowbreak{}a065b956\allowbreak{}7891e45c\allowbreak{}79542d73\allowbreak{}57} \\
\bottomrule
\end{tabular}
\end{table}

USDC amounts are converted using 6 decimal places; MOG, UNIBOT, and WETH use 18 decimal places. The historical TRU implementation used in external validation has 8 decimal places. Calculations retain raw integers, while displayed amounts use the corresponding token's decimal places.

Historical control records support the entry-point closure findings for Socket, LiFi, Maestro, and Dolomite. For Maestro, post-upgrade call dispatch evidence confirms that the old entry point was disabled. Unibot uses the public pause time as the remediation reference, while Sushi's interface remediation time is recorded as unknown. Complete block hashes and source mappings for each case are included in the case evidence index in Appendix~\ref{app:D.3}.

\subsection{Scope of Historical Data Cross-Checks}\label{app:A.3}

Allowance and balance are obtained from the same historical block and together form a state record. All 2,406 applicable state records in the six-case design set are complete, including 470 cross-checked against an independent source. A total of 537 transaction evidence packages support review of operation execution and fund destinations.

\begin{figure}[tp]
\centering
\includegraphics[width=\textwidth,height=0.82\textheight,keepaspectratio]{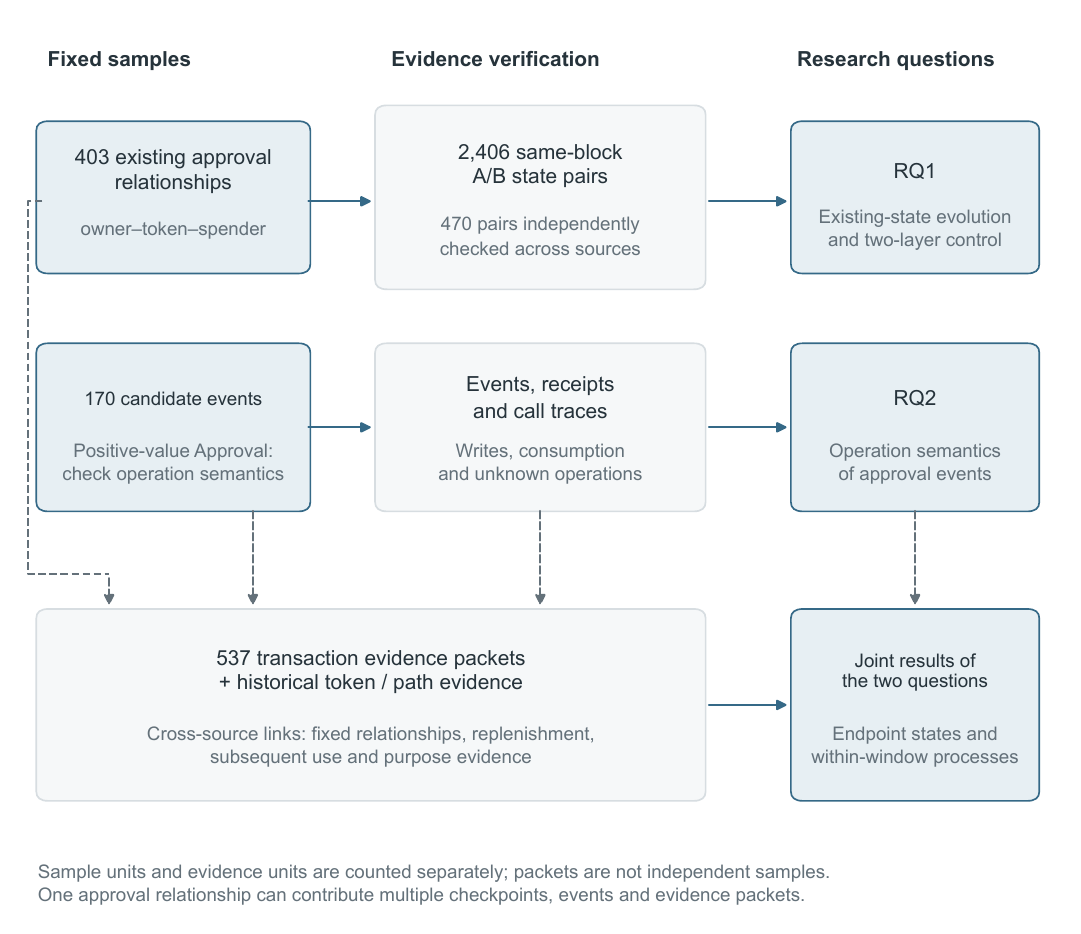}
\renewcommand{\thefigure}{A1}
\caption{Evidence Links in the Design Set. The 2,406 allowance--balance records describe historical states; 470 independently checked records provide cross-checks; and 537 transaction evidence packages explain actual operations. States and episodes from external validation are reported separately.}\label{fig:A1}
\end{figure}

External validation obtained 106 main-analysis state records and 80 disclosure-lower-bound sensitivity records, with 11 and 8 records sampled for cross-checking, respectively. Another 32 records at major state-transition boundaries were also cross-checked. Reproduction materials preserve query ranges, deduplication results, raw responses, and file checksums so that aggregate results can be traced to their sources. Figure~\ref{fig:A1} shows the evidence links from state queries to execution verification in the design set.

\subsection{Additional Views of Control Scope and Allowance Cleanup}\label{app:A.4}

\begin{figure}[tp]
\centering
\includegraphics[width=\textwidth,height=0.82\textheight,keepaspectratio]{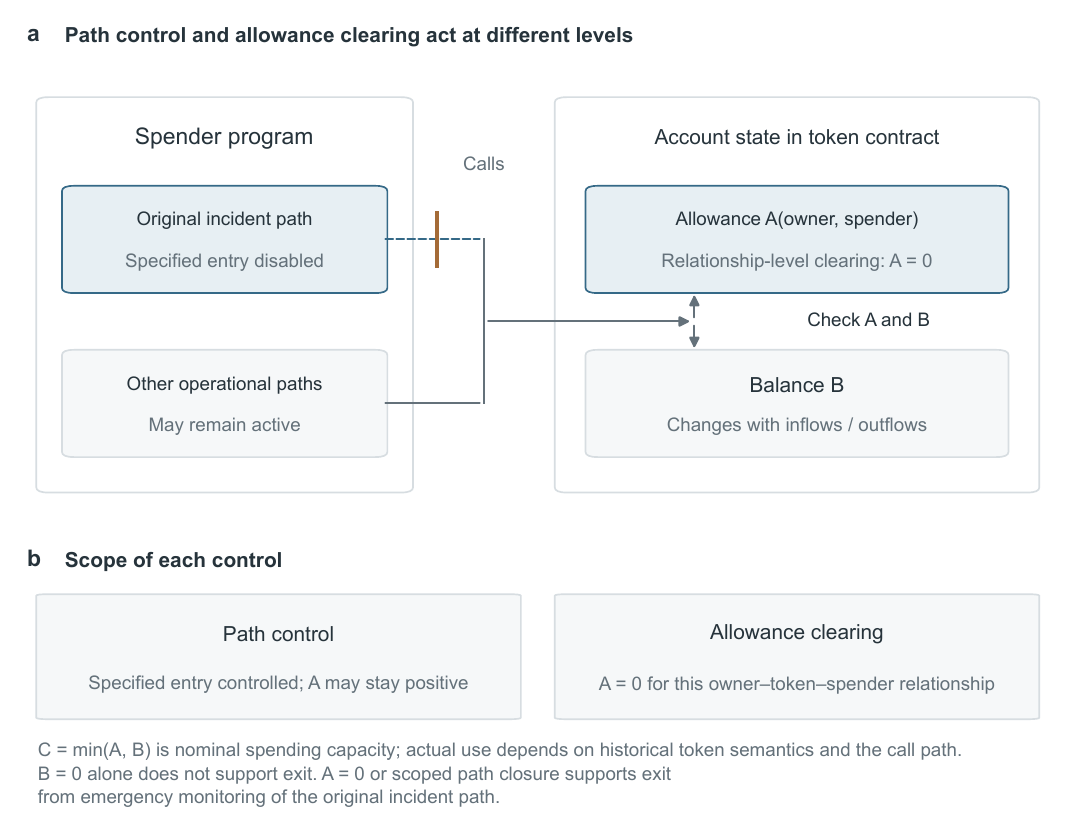}
\renewcommand{\thefigure}{A2}
\caption{Approval relationships and the targets of the two types of remediation. Software-side controls apply to the original incident path; allowance changes apply to individual approval relationships.}\label{fig:A2}
\end{figure}

Applications may use proxy contracts to separate an entry point from the logic actually executed. Proxies can support upgrades and functions such as code sharing \cite{ref11,ref12}. We identify software-side control scope from the historical entry point and implementation, recording address changes separately. Figure~\ref{fig:A2} shows the distinct targets of program controls and account-level allowance changes.

\begin{figure}[tp]
\centering
\includegraphics[width=\textwidth,height=0.82\textheight,keepaspectratio]{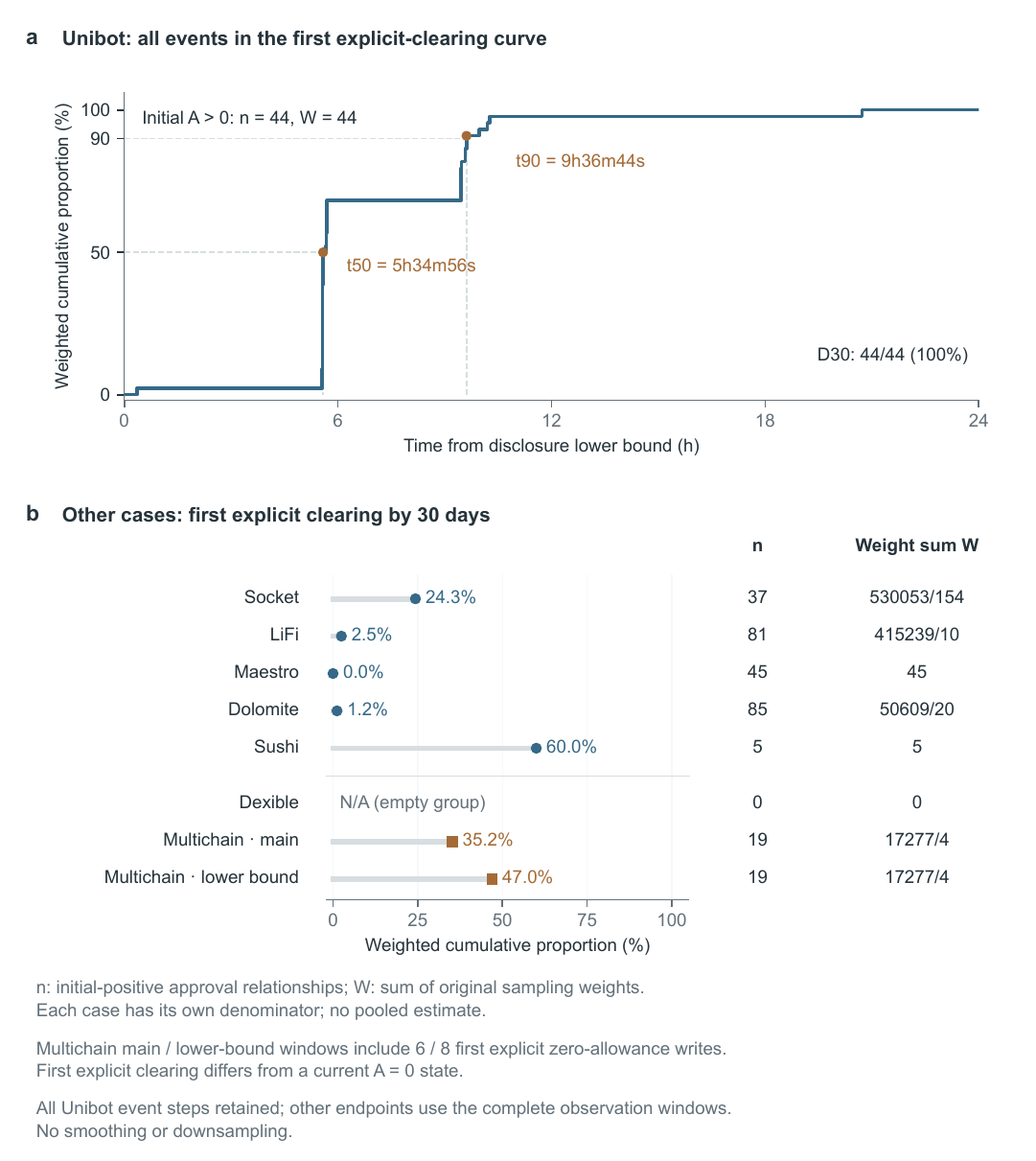}
\renewcommand{\thefigure}{A3}
\caption{First explicit setting of allowances to zero among relationships with initially positive allowances. The curves and endpoint proportions use stratified weights to measure cumulative progress in explicitly setting allowances to zero. Allowances consumed to zero and current zero allowances (A=0) are recorded separately. For Multichain, allowances in 6 relationships were explicitly set to zero after the primary observation start; another 2 were within the disclosure interval. Dexible's initially positive allowance group is empty, so times to specified cleanup proportions do not apply.}\label{fig:A3}
\end{figure}

Figure~\ref{fig:A3} retains the full clearance trajectory and case-level weighted endpoints. Each case uses its own initially positive allowance group, including an empty group for Dexible.

\section{Verification of Approval Operations and Selected Fund-Flow Episodes}\label{app:B}

\subsection{Identifying Operations and Transfer Purposes}\label{app:B.1}

Approval-related logs are classified as setting a positive allowance, spending against an existing allowance, or undetermined. When the exact allowance before an operation can be recovered, we further distinguish increases, repeated settings, reductions, and reapproval after the allowance was set to zero. This links permission changes recorded in logs to the operations that actually occurred.

The call that executed a transfer, the source of its permission, and its purpose are verified separately. The transaction initiator and fund destination determine whether a transfer is recorded as account-initiated or third-party-directed. Specific purposes, such as normal business activity, attack, or rescue, are determined using corresponding execution evidence and public materials. Eight selected episodes link incoming funds and subsequent spending by the application under study to the same approval relationship. U2 and U3 also have support from a contemporaneous public alert \cite{ref10}.

Token-specific reconstruction follows the historical implementations. TRU's post-spending allowance is recovered directly from logs, without subtracting Transfer amounts again; USDC allowance consumption must be linked to actual calls. WETH balances also require Deposit and Withdrawal events, and WETH's exact maximum allowance does not decrease during spending. MOG uses a specifically verified balance model.

\subsection{Transaction Mapping for Eight Fund-Flow Episodes}\label{app:B.2}

Table~\ref{tab:B1} lists eight episodes across three Unibot relationships and two Sushi relationships. Each episode connects an incoming transaction to a subsequent spending transaction, checking intervening allowance settings and execution order. Multiple episodes can occur in the same wallet, so the eight records represent five distinct relationships.

\begin{table}[!htbp]
\centering\footnotesize
\renewcommand{\thetable}{B1}
\setlength{\tabcolsep}{4pt}
\renewcommand{\arraystretch}{1.16}
\caption{Selected Fund-Flow Episodes and Transaction Locations. ``Block number\#index'' identifies a transaction's block and its position within that block; same-block episodes are ordered by index. A star indicates support from a contemporaneous alert.}\label{tab:B1}
\begin{tabular}{>{\raggedright\arraybackslash}p{\dimexpr 0.090000\textwidth-2\tabcolsep\relax}>{\raggedright\arraybackslash}p{\dimexpr 0.200000\textwidth-2\tabcolsep\relax}>{\raggedright\arraybackslash}p{\dimexpr 0.200000\textwidth-2\tabcolsep\relax}>{\raggedright\arraybackslash}p{\dimexpr 0.200000\textwidth-2\tabcolsep\relax}>{\raggedright\arraybackslash}p{\dimexpr 0.310000\textwidth-2\tabcolsep\relax}}
\toprule
\textbf{Episode} & \textbf{Incoming block\#index} & \textbf{Spending block\#index} & \textbf{Interval} & \textbf{Transaction records} \\
\midrule
U1 & 18469019\#79 & 18469020\#0 & 12 seconds & \href{https://etherscan.io/tx/0x4886cb8b9e4281a069beca5b02dafa542a08f99f37bf6fbe730d6b9757b11e25}{Incoming} / \href{https://etherscan.io/tx/0xa846c5189f07df30c53172b032e3cb295f8c0fb6625d910ef52f9ca0ef32dd5a}{Spending} \\
U2\(\star\) & 18468914\#15 & 18468918\#29 & 60 seconds & \href{https://etherscan.io/tx/0x6b96da67e0ac8e520feb91b84eba44633729a9f4e07b86f844bd4affff6cc011}{Incoming} / \href{https://etherscan.io/tx/0x875acc67d001505134e6e76fa322fbc86e59e6e4e1b685a69ccc6ae854056d52}{Spending} \\
U3\(\star\) & 18468922\#38 & 18468929\#0 & 84 seconds & \href{https://etherscan.io/tx/0x3c7d7c9684e84dc22f8ec004c10831f5787c52eb8b33c0bb349fffd0cb6dfe80}{Incoming} / \href{https://etherscan.io/tx/0xa32c32f8b1d81bf1ee79adb072bcfbdc7fd9fa3fbd5219b3439c2d51d05f9d99}{Spending} \\
U4 & 18468798\#21 & 18468809\#13 & 132 seconds & \href{https://etherscan.io/tx/0x29f5601dae30218fa3e4fd96adc9d97d61f7807ed9dc1941743a2babe7d3a957}{Incoming} / \href{https://etherscan.io/tx/0x552c53deba7a80a7d4331ddb0728a6d53593e8f62eb824ef30155b32c563b78c}{Spending} \\
S1 & 17010646\#140 & 17010648\#14 & 24 seconds & \href{https://etherscan.io/tx/0x5b57a290d9e3fe9a0064a0d1b6c3b12396eaf220249775a0cba48ae281f50d9c}{Incoming} / \href{https://etherscan.io/tx/0x0ea3b80cb14fc69058daccdd68de79a70475aec9208b7fd573ae5e85c075a2cf}{Spending} \\
S2 & 17013640\#0 & 17013640\#1 & Same block, ordered by index & \href{https://etherscan.io/tx/0x3c08dc41e49686e537563872065961cd4f0d81a3ddffce0aabc3726874cbed38}{Incoming} / \href{https://etherscan.io/tx/0xd1d160efab56da105f5fde0b021f296b8714b731b287abc6286567b873cd780e}{Spending} \\
S3 & 17014353\#1 & 17014353\#2 & Same block, ordered by index & \href{https://etherscan.io/tx/0xaff34c75e79197e7971d0a7597dcf9f73ecc71c80b3970a594919ffe54579f03}{Incoming} / \href{https://etherscan.io/tx/0x6c045f46828e4a629fa819f584239d6b6e70ffd91955a4c2ccc01cad40c3fd6b}{Spending} \\
S4 & 17081172\#117 & 17081173\#10 & 12 seconds & \href{https://etherscan.io/tx/0x6f69a6cedb8a06fc5f96d0b90c7089ec003bf5f1a8a4fcf9ba56455c9bfeaa14}{Incoming} / \href{https://etherscan.io/tx/0xbc8ea0cec422fd903cb54e0400350b904f0e752499618817383e47432f9a6320}{Spending} \\
\bottomrule
\end{tabular}
\end{table}

U2 and U3 occurred in the same wallet. At 08:26:59 on 2023-10-31, the wallet bought back UNIBOT; approximately 29.9169 UNIBOT was extracted 60 seconds later. It bought back UNIBOT again at 08:28:47, and approximately 11.0842 UNIBOT was extracted 84 seconds later. Both spending events used permissions retained from before the buybacks, with zero intervening positive allowance rewrites between each buyback and extraction. The allowance for this relationship was explicitly set to zero at 10:52:59. The addresses, amounts, and sequence in the contemporaneous PeckShield alert match both records \cite{ref10}.

S1, S2, and S3 also occurred in the same wallet. Sushi's public response materials describe ongoing rescue efforts for future incoming funds \cite{ref14,ref15}, providing context for interpreting subsequent activity in this case. Each episode's purpose is labeled according to its transaction evidence. The links in Table~\ref{tab:B1} contain complete transaction identifiers. The selected-episode transaction list in Appendix~\ref{app:D.3} also provides accounts, tokens, applications, block hashes, UTC times, and raw integer amounts.

\begin{figure}[tp]
\centering
\includegraphics[width=\textwidth,height=0.82\textheight,keepaspectratio]{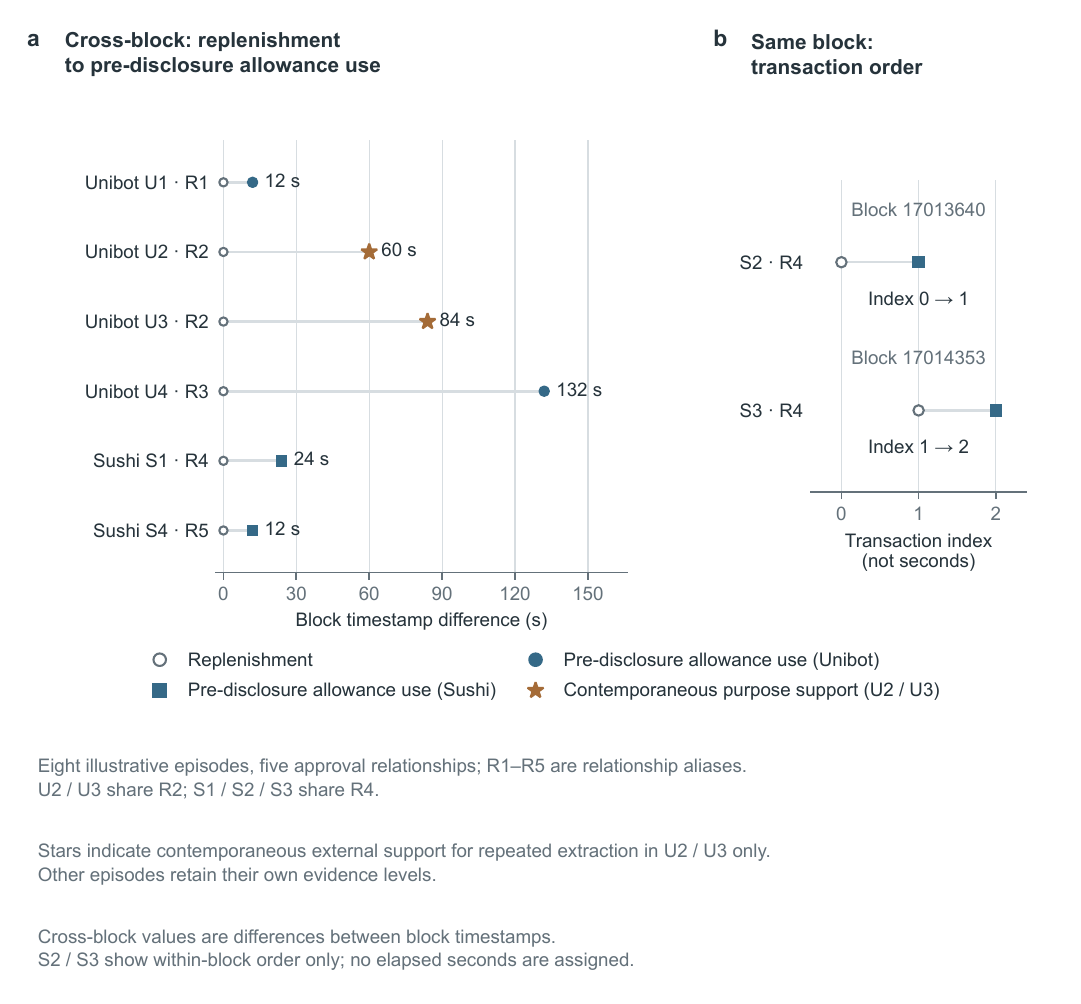}
\renewcommand{\thefigure}{A4}
\caption{Eight Fund-Flow Episodes in the Design Set. Cross-block episodes show elapsed times, while same-block episodes show transaction order; episodes belonging to the same relationship share an identifier. U2 and U3 have support from a contemporaneous alert. The design set and external validation are reported separately, with purpose labels retained for each episode.}\label{fig:A4}
\end{figure}

\subsection{Episodes with Incompletely Recovered Allowances}\label{app:B.3}

The 14 episodes with unknown allowances at replenishment in Section~\ref{sec:4.3} are concentrated in one LiFi relationship. Its permission history alternates between allowance settings and consumption. Available operation records do not fully cover the exact allowances before some transactions, so the allowances at all 14 replenishments are recorded as unknown. Actual spending by the application under study and historical state points verified through queries each have supporting evidence.

Evidence status is recorded separately for allowances, entry-point closure, and transfer purposes. The statistics retain these 14 episodes and their unknown allowances, while verified calls describe subsequent fund activity.

\section{Sampling, Statistical Calculations, and Robustness Analysis}\label{app:C}

\subsection{Sample Selection and Calculation of Counts and Intervals}\label{app:C.1}

\textbf{Stratification and sampling.} To cover both older approvals and recently active approvals, fixed relationships are grouped by the last occurrence of a relevant positive allowance log. Those with such logs within the last 90 days enter the recently active stratum (recent); the others enter the earlier-activity stratum (historical). Each case includes at most 100 relationships, with all relationships included when the sampling frame is below this limit. Larger frames receive proportional allocations, with each nonempty stratum guaranteed at least min(20,\(N_{h}\)) relationships, where \(N_{h}\) is the total number of objects in that stratum.

Post-incident positive allowance log candidates are grouped into 0--1, 1--7, and 7--30 days after disclosure, with at most 50 selected per case. Each nonempty stratum first receives min(5,\(N_{h}\)) places; remaining places are allocated according to each stratum's remaining capacity. External validation follows the same relationship-sampling method, with a cap of 20 per case and a minimum of min(4,\(N_{h}\)) for each nonempty stratum.

Sample lists are fixed before formal queries for balances, allowances, and subsequent outcomes. The design set uses the seed formal-v3-20261026, and external validation uses external-B-v1-20261103. Canonical relationship or event identifiers are concatenated with the seed, and selection follows the order of their SHA-256 values. Dexible's frame contains 1 relationship, which is included in full. Multichain selects 20 of 4,573 relationships: 4 of 513 in the earlier-activity stratum and 16 of 4060 in the recently active stratum.

\begin{table}[!htbp]
\centering\footnotesize
\renewcommand{\thetable}{C1}
\setlength{\tabcolsep}{4pt}
\renewcommand{\arraystretch}{1.16}
\caption{Stratum Sizes and Selected Counts. Each cell reports ``total objects N / sample size n''; 0/0 denotes an empty stratum. RQ1 concerns fixed relationships, and RQ2 concerns post-incident candidate events.}\label{tab:C1}
\begin{tabular}{>{\raggedright\arraybackslash}p{\dimexpr 0.150000\textwidth-2\tabcolsep\relax}>{\raggedright\arraybackslash}p{\dimexpr 0.170000\textwidth-2\tabcolsep\relax}>{\raggedright\arraybackslash}p{\dimexpr 0.170000\textwidth-2\tabcolsep\relax}>{\raggedright\arraybackslash}p{\dimexpr 0.170000\textwidth-2\tabcolsep\relax}>{\raggedright\arraybackslash}p{\dimexpr 0.170000\textwidth-2\tabcolsep\relax}>{\raggedright\arraybackslash}p{\dimexpr 0.170000\textwidth-2\tabcolsep\relax}}
\toprule
\textbf{Case} & \textbf{RQ1 earlier-activity stratum} & \textbf{RQ1 recently active stratum} & \textbf{RQ2 0--1 days} & \textbf{RQ2 1--7 days} & \textbf{RQ2 7--30 days} \\
\midrule
Socket & 5196/56 & 4102/44 & 24/5 & 309/11 & 1421/34 \\
LiFi & 25435/50 & 25815/50 & 48/5 & 1225/10 & 7050/35 \\
Maestro & 0/0 & 47/47 & 7/7 & 20/17 & 32/26 \\
Unibot & 0/0 & 44/44 & 20/20 & 0/0 & 0/0 \\
Dolomite & 2977/100 & 0/0 & 0/0 & 0/0 & 0/0 \\
Sushi RP2 & 0/0 & 12/12 & 0/0 & 0/0 & 0/0 \\
\bottomrule
\end{tabular}
\end{table}

\textbf{Weighted estimation.} Sampling fractions differ across strata, so each sampled object represents a different number of objects in the frame. For stratum h, let \(N_{h}\) denote its total size, \(n_{h}\) its sample size, and \(y_{h}\) the number of sampled objects meeting the condition being counted. Each sampled object then has weight \(N_h/n_h\), and the estimated count is \((N_h/n_h)y_h\). Summing these estimates across strata gives the count estimate for the corresponding sampling frame. Dividing by the corresponding denominator for the sampling frame or prespecified subgroup gives the proportion estimate. Weights for empty strata are recorded as not applicable.

Observed counts and weighted estimates are reported together. Zero-setting curves use descriptive weighted proportions calculated under each case's original sampling design. For Multichain, zero-setting within the main window and cumulative zero-setting including disclosure-interval events share the same denominator. This denominator is the initially positive allowance group established before the disclosure lower bound.

\textbf{Estimation uncertainty.} Interval calculations approximate the prespecified hash-based selection as equal-probability sampling without replacement within each stratum. Count intervals use \(\alpha=0.05\). With \(H_{s}\) denoting the number of strata requiring sampling, each stratum receives an error allocation of \(\alpha/H_s\), with \(\alpha/(2H_s)\) in each tail. Equal-tailed hypergeometric inversion yields a population count interval compatible with the observed count in each stratum.

Fully enumerated strata use their observed counts and ranges compatible with unknown records directly, with zero sampling error. When all strata are fully enumerated, their counts and compatible ranges are summed directly. Summing stratum lower bounds and upper bounds gives the two endpoints of the overall count interval. Dividing by the known sampling-frame size gives the corresponding proportion interval.

Unknown records always remain in their original denominators. For the compatible lower bound, all unknown records are treated as not meeting the condition; for the compatible upper bound, all are treated as meeting it. This presents both confirmed counts and the range permitted by the available evidence.

For prespecified subgroup proportions whose denominators are also estimated from the sample, the numerator and denominator each receive \(\alpha/2\). Count intervals are then constructed using the stratum allocation rules above. The proportion's lower bound is the numerator's lower bound divided by the denominator's upper bound. When the denominator's lower bound is positive, the proportion's upper bound is the numerator's upper bound divided by the denominator's lower bound, capped at 1. If the denominator's lower bound is zero and its upper bound is positive, the proportion's upper bound depends on the numerator's upper bound. It is 0 when the numerator's upper bound is zero and 1 otherwise. If the denominator's upper bound is zero, the proportion is recorded as not applicable. Calculations preserve the constraints that the numerator cannot exceed the denominator and that a valid proportion requires a positive denominator.

These intervals cover individual named metrics and their strata, with separate calculations for protocols, time points, and outcome categories. Original fractional weights and sampling configurations are listed in the accompanying evidence index in Appendix~\ref{app:D.3}.

\subsection{Allowance Units and Sources of Permission at Day 30}\label{app:C.2}

An allowance represents an application's retained permission to transfer tokens. Raw allowances and balances are stored as integers and converted using the token's decimal places only for display. We label A\(\geq\)\(2^{255}\) as a ``near-unlimited allowance'' and an allowance exactly equal to \(2^{256}\)-1 as the maximum value (MAX). Whether each use reduces the allowance is determined from the token's code at that time.

Table~\ref{tab:C2} links relationships with both positive allowances and positive balances at day 30 to their earlier permission records. It checks whether these permissions were retained from before disclosure or set afterward.

We use the checkpoints mapped to the disclosure lower bounds in the six design cases. At these checkpoints, 269 of the 297 relationships with positive allowances met the near-unlimited threshold. Of these, 141 were exactly at the maximum allowance. At day 30, 90 of the 97 relationships with both positive allowances and balances still met that threshold. Residual nominal spending capacity across the three USDC cases totaled 2,559.405124 USDC at that time. This total sums the smaller of allowance and balance for each relationship.

\begin{table}[!htbp]
\centering\footnotesize
\renewcommand{\thetable}{C2}
\setlength{\tabcolsep}{4pt}
\renewcommand{\arraystretch}{1.16}
\caption{Relationships with Positive Allowances and Balances at Day 30 and Their Permission Histories. A denotes the transfer allowance; C>0 means that both allowance and balance are positive. All counts in the table are approval relationships.}\label{tab:C2}
\begin{tabular}{>{\raggedright\arraybackslash}p{\dimexpr 0.170000\textwidth-2\tabcolsep\relax}>{\raggedright\arraybackslash}p{\dimexpr 0.150000\textwidth-2\tabcolsep\relax}>{\raggedright\arraybackslash}p{\dimexpr 0.170000\textwidth-2\tabcolsep\relax}>{\raggedright\arraybackslash}p{\dimexpr 0.260000\textwidth-2\tabcolsep\relax}>{\raggedright\arraybackslash}p{\dimexpr 0.250000\textwidth-2\tabcolsep\relax}}
\toprule
\textbf{Case} & \textbf{C>0 at day 30} & \textbf{A>0 before disclosure} & \textbf{Positive allowance set after disclosure} & \textbf{Allowance explicitly set to zero after disclosure} \\
\midrule
Socket & 9 & 9 & 0 & 0 \\
LiFi & 46 & 46 & 0 & 0 \\
Maestro & 33 & 32 & 1 & 0 \\
Unibot & 0 & 0 & 0 & 0 \\
Dolomite & 9 & 9 & 0 & 0 \\
Sushi RP2 & 0 & 0 & 0 & 0 \\
Total & 97 & 96 & 1 & 0 \\
\bottomrule
\end{tabular}
\end{table}

Of the 97 relationships, 96 had positive allowances before disclosure that persisted through the end of the window. No positive allowance-setting or explicit zero-setting operations occurred during this period. The other relationship had a zero allowance before disclosure and received a positive setting later. We observed no allowance-setting operations in the corresponding disclosure-boundary block. No sequence of explicitly setting the allowance to zero and then setting a positive allowance occurred within the window.

The transaction-level state before this positive allowance setting is recorded as unknown, and the event is reported as a permission setting within the observation window. Complete relationship identifiers, log coverage, and event sources are included in the day-30 permission history list.

\subsection{Effects of Very Small Amounts and Prior Exposure}\label{app:C.3}

\textbf{Amount thresholds.} To check whether fund activity mainly involved very small amounts, we filter post-inflow balances at thresholds of 1, 10, and 100 tokens. This filtering applies to the original episodes in which balances changed from zero to positive. For spending by the application under study, we also filter the amount of each call. Each case uses its token's native units, and episode start and end definitions follow the original zero-to-positive records.

\begin{table}[!htbp]
\centering\footnotesize
\renewcommand{\thetable}{C3}
\setlength{\tabcolsep}{4pt}
\renewcommand{\arraystretch}{1.16}
\caption{Relationship Counts after Amount Filtering. ``Replenishment'' means that an account balance changed from zero to positive. ``Spending'' means that the application under study actually exercised the transfer permission for that relationship. Counts at the three thresholds are listed in order.}\label{tab:C3}
\begin{tabular}{>{\raggedright\arraybackslash}p{\dimexpr 0.130000\textwidth-2\tabcolsep\relax}>{\raggedright\arraybackslash}p{\dimexpr 0.100000\textwidth-2\tabcolsep\relax}>{\raggedright\arraybackslash}p{\dimexpr 0.150000\textwidth-2\tabcolsep\relax}>{\raggedright\arraybackslash}p{\dimexpr 0.230000\textwidth-2\tabcolsep\relax}>{\raggedright\arraybackslash}p{\dimexpr 0.150000\textwidth-2\tabcolsep\relax}>{\raggedright\arraybackslash}p{\dimexpr 0.240000\textwidth-2\tabcolsep\relax}}
\toprule
\textbf{Case} & \textbf{Sample size n} & \textbf{Original replenishment relationships} & \textbf{Replenishment relationships at thresholds 1/10/100} & \textbf{Original application-spending relationships} & \textbf{Spending relationships at thresholds 1/10/100} \\
\midrule
Socket & 100 & 17 & 17/17/16 & 5 & 5/5/5 \\
LiFi & 100 & 12 & 12/12/10 & 6 & 6/6/5 \\
Maestro & 47 & 12 & 12/12/12 & 20 & 20/20/20 \\
Unibot & 44 & 34 & 28/19/10 & 15 & 4/3/1 \\
Dolomite & 100 & 5 & 5/5/4 & 0 & 0/0/0 \\
Sushi RP2 & 12 & 5 & 0/0/0 & 3 & 0/0/0 \\
\bottomrule
\end{tabular}
\end{table}

For Sushi, the number of replenishment relationships meeting the filter is 0 from the 1 WETH threshold onward. For Unibot, the number of relationships with use of an existing allowance after replenishment is 0 at the 100 UNIBOT threshold. These changes show that activity counts vary with the token amount range of interest.

\textbf{Prior exposure.} Relationships investigated or used for technical test reads earlier in the research are labeled as having ``prior exposure.'' We exclude these relationships from the original fixed sample and apply the same definitions to the remaining relationships. This checks how much the results depend on objects encountered in the earlier work.

After excluding 25 prior-exposure relationships from Unibot, replenishment relationships decrease from 34 to 12, and application-spending relationships decrease from 15 to 2. Relationships with third-party spending after replenishment decrease from 3 to 0. The latter observation is therefore concentrated in the prior-exposure group.

\begin{table}[!htbp]
\centering\footnotesize
\renewcommand{\thetable}{C4}
\setlength{\tabcolsep}{4pt}
\renewcommand{\arraystretch}{1.16}
\caption{Observed Counts after Excluding Prior-Exposure Relationships}\label{tab:C4}
\begin{tabular}{>{\raggedright\arraybackslash}p{\dimexpr 0.170000\textwidth-2\tabcolsep\relax}>{\raggedright\arraybackslash}p{\dimexpr 0.140000\textwidth-2\tabcolsep\relax}>{\raggedright\arraybackslash}p{\dimexpr 0.140000\textwidth-2\tabcolsep\relax}>{\raggedright\arraybackslash}p{\dimexpr 0.170000\textwidth-2\tabcolsep\relax}>{\raggedright\arraybackslash}p{\dimexpr 0.180000\textwidth-2\tabcolsep\relax}>{\raggedright\arraybackslash}p{\dimexpr 0.200000\textwidth-2\tabcolsep\relax}}
\toprule
\textbf{Case} & \textbf{Original sample size} & \textbf{Sample size after exclusion} & \textbf{Replenishment relationships} & \textbf{Application-spending relationships} & \textbf{Relationships with spending after replenishment} \\
\midrule
Socket & 100 & 100 & 17 & 5 & 2 \\
LiFi & 100 & 100 & 12 & 6 & 4 \\
Maestro & 47 & 26 & 7 & 12 & 5 \\
Unibot & 44 & 19 & 12 & 2 & 0 \\
Dolomite & 100 & 100 & 5 & 0 & 0 \\
Sushi RP2 & 12 & 11 & 5 & 3 & 2 \\
\bottomrule
\end{tabular}
\end{table}

Table~\ref{tab:C4} reports observed relationship counts. Post-exclusion frame proportions retain the original weights and are estimated within the subgroup of remaining relationships. Complete numerators, denominators, and 95\% intervals are provided in the accompanying sensitivity estimates table.

\textbf{Choice of disclosure time.} Sushi is also checked using an alternative notification time of 2023-04-09 at 04:05:22.724 UTC. Among the same 12 relationships, 5 still show returning funds, 3 show use of an existing allowance, and 2 show such use after funds return. Individual state records at the disclosure and day-30 boundaries are preserved in the time-boundary and sensitivity counts table.

\subsection{Dependence of Results on Case Composition}\label{app:C.4}

We remove one protocol at a time to examine changes in the average results for the remaining protocols. Proportions are first calculated within each case using the original weights, then averaged with equal weight across included protocols. The full result includes 6 cases, and each leave-one-out result includes 5. This comparison describes the sensitivity of aggregate values to case composition.

\begin{table}[!htbp]
\centering\footnotesize
\renewcommand{\thetable}{C5}
\setlength{\tabcolsep}{4pt}
\renewcommand{\arraystretch}{1.16}
\caption{Protocol-Average Proportions with All Cases and with Each Case Removed (\%)}\label{tab:C5}
\begin{tabular}{>{\raggedright\arraybackslash}p{\dimexpr 0.170000\textwidth-2\tabcolsep\relax}>{\raggedright\arraybackslash}p{\dimexpr 0.150000\textwidth-2\tabcolsep\relax}>{\raggedright\arraybackslash}p{\dimexpr 0.170000\textwidth-2\tabcolsep\relax}>{\raggedright\arraybackslash}p{\dimexpr 0.160000\textwidth-2\tabcolsep\relax}>{\raggedright\arraybackslash}p{\dimexpr 0.170000\textwidth-2\tabcolsep\relax}>{\raggedright\arraybackslash}p{\dimexpr 0.180000\textwidth-2\tabcolsep\relax}}
\toprule
\textbf{Excluded case} & \textbf{Positive allowance at day 30} & \textbf{Positive capacity at day 30} & \textbf{Replenishment} & \textbf{Application spending} & \textbf{Spending after replenishment} \\
\midrule
All retained (6 cases) & 51.12 & 22.37 & 29.75 & 18.78 & 8.11 \\
Socket & 55.74 & 25.05 & 32.30 & 21.54 & 9.33 \\
LiFi & 45.74 & 17.64 & 33.29 & 21.33 & 8.93 \\
Maestro & 41.34 & 12.81 & 30.59 & 14.03 & 5.90 \\
Unibot & 61.34 & 26.85 & 20.24 & 15.72 & 8.37 \\
Dolomite & 44.54 & 25.05 & 34.70 & 22.54 & 9.73 \\
Sushi RP2 & 58.01 & 26.85 & 27.36 & 17.54 & 6.40 \\
\bottomrule
\end{tabular}
\end{table}

The columns examine permissions and spending capacity at the end of the window, and fund activity during the month. The descriptive leave-one-out proportions show how each metric varies with case composition. Detailed numerators and denominators are provided in the protocol-composition sensitivity table.

\subsection{Monitoring Rules, Evidence Timing, and State Updates}\label{app:C.5}

\textbf{Comparison targets and decision conditions.} Monitoring targets are specific approval relationships and their corresponding original incident paths. Under the same information scenario, all rules read the same state and control evidence to decide whether a relationship remains on the emergency monitoring list.

The balance rule retains a relationship when balance B is positive or unknown; the capacity rule retains it when nominal spending capacity C is positive or undetermined. If either allowance A or balance B is confirmed to be zero, C is determined to be zero, and the capacity rule exits on that basis. The allowance rule R0 retains a relationship when A is positive or unknown and exits when A is confirmed to be zero.

The allowance-and-remediation rule R1 accepts two grounds for exit: the allowance is already zero, or evidence establishes closure of the original affected entry point. This closure evidence must be available at the decision time and applicable to the relationship. The full evidence review F examines whether available materials support reasons to retain or exit beyond those used by R1. Positive allowance settings and verified control changes trigger reassessment; initial inclusion, continued retention, and reentry after exit are recorded separately.

\textbf{Observation times.} The main comparison uses four common checkpoints: the disclosure boundary and days 1, 7, and 30 after disclosure. Fund-flow episode analysis also uses preserved, verifiable state updates. The decision before spending uses the latest record from a block strictly earlier than the spending block. A relationship retained at all four checkpoints contributes four ``relationship--checkpoint'' retention decisions. The evaluation metric is the change in the list at these checkpoints and block-state updates.

\textbf{When evidence becomes available.} The retrospective interpretation scenario H uses control interpretations verified as of now and associates each measure with its historical effective time. The contemporaneously available evidence scenario O determines which interpretations can be used from verifiable records of contemporaneous availability. The availability status of other interpretations is recorded as unknown. The two scenarios separately represent historical control facts and the materials on which a monitor could have based decisions at the time.

Additional hypothetical scenarios assume delays of 0, 1, and 24 hours in producing an interpretation. Delay starts at the end of the actual control block. Once the specified time has elapsed, the decision is updated at the first available block record that meets the timing condition. Each decision uses the state and control evidence permitted by its information scenario; subsequent purpose interpretations are used to verify results.

\textbf{Retention period and reentry.} The monitoring retention period runs from a relationship's entry onto the monitoring list until grounds for its exit become available. Path controls, allowance consumption to zero, and the availability of closure evidence can each determine its endpoint. Durations are reported for intervals with continuous evidence; verified time points or transition-time ranges are reported for intervals with discrete evidence. Main tables consistently count relationship--checkpoints.

In the design set's retrospective interpretation scenario, R0 has 3 reentries after exit, all from 1 relationship. Relationships absent from the initial list that enter later for the first time are counted separately. These changes document how permission and evidence updates alter monitoring decisions.

\textbf{Tracing results to evidence.} Remediation records include the controlled address, function entry point, applicable version, effective block, evidence source, and actions still required of accounts. They separately preserve the times when measures took effect and when interpretations became available. Monitoring decisions link specific relationships, observation blocks, and the evidence used. Fund-flow episodes link the transaction order of incoming funds and spending, the application that actually transferred funds, and the source of permission. Purpose evidence is linked to specific transactions.

\begin{figure}[tp]
\centering
\includegraphics[width=\textwidth,height=0.82\textheight,keepaspectratio]{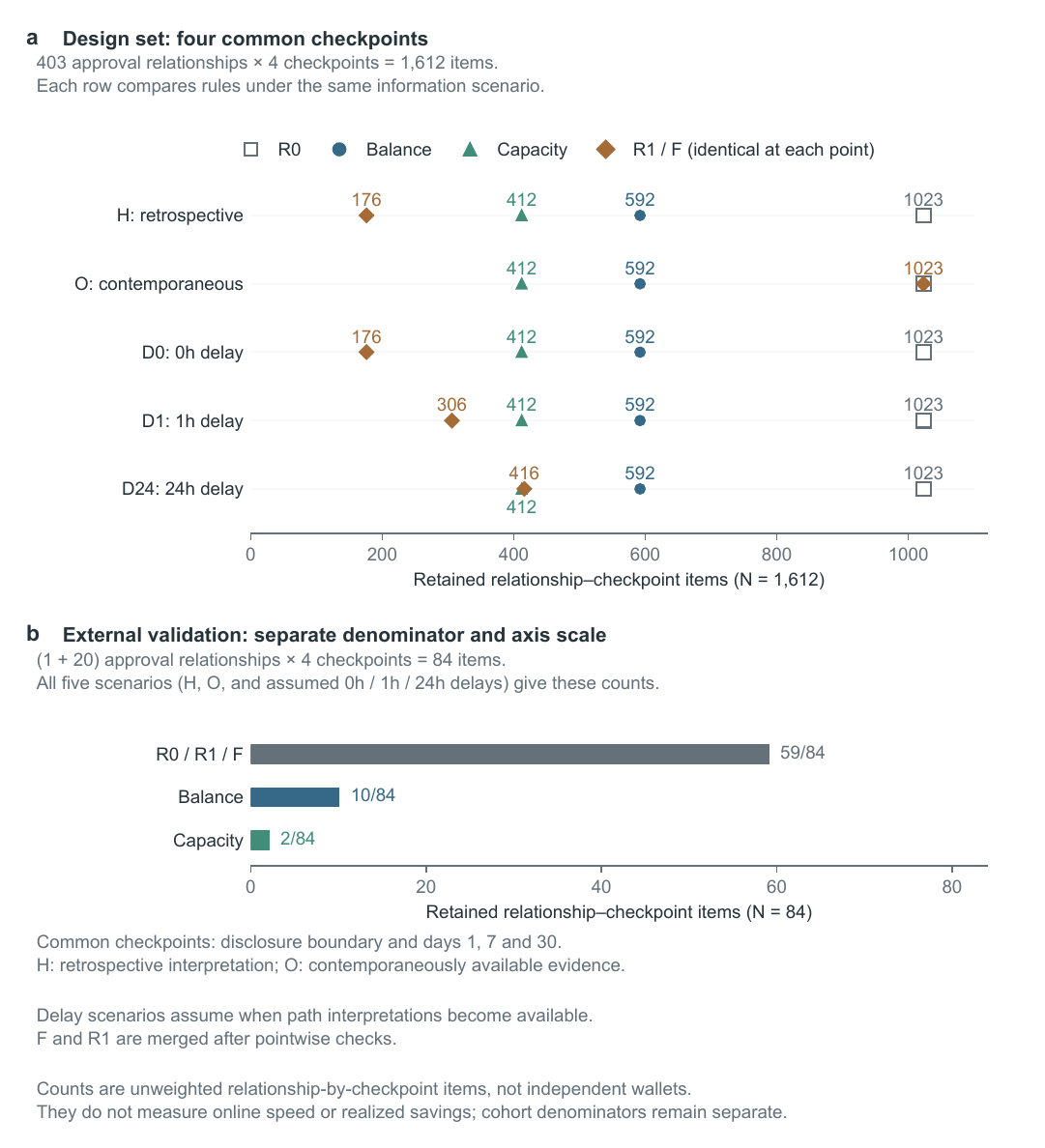}
\renewcommand{\thefigure}{A5}
\caption{Dependence of monitoring outcomes on evidence conditions. Across six cases, R1 and F retain 176 items under retrospectively approved interpretations (H). They retain 1,023 when using only evidence confirmed to have been available at the time (O). Hypothetical interpretation delays of 1 hour and 24 hours yield 306 and 416 retained items, respectively. For the two external cases, R0, R1, and F each retain 59 items in all five scenarios. The two groups have denominators of 1,612 and 84 relationship--checkpoints, respectively. D0, D1, and D24 in the figure denote interpretation delays of 0, 1, and 24 hours, respectively.}\label{fig:A5}
\end{figure}

\section{Analysis Toolkit and Reproduction Validation}\label{app:D}

\subsection{Analysis Tasks and Inputs}\label{app:D.1}

The Authorization Recovery Toolkit organizes preserved on-chain records into verifiable approval histories and fund-flow episodes and recomputes the study summaries reported in this paper. The toolkit consists of two connected components.

The first component processes transactions, execution receipts, call records, and token events to identify allowance settings and consumption. It reconstructs balances and allowances in transaction order and identifies the application that actually exercised the permission. Previously checked materials provide the token's operating rules at the time, code version, scope of remediation, and reviewed purpose labels. Thus, automatically processed execution facts and manually reviewed semantic interpretations each have identified sources in the inputs.

The second component reads reviewed relationships, checkpoints, and fund-flow episodes to calculate progress toward first zero-setting, day-30 states, episode groups, and monitoring decisions. Aggregate results retain source mappings so that table counts can be traced to individual records. Raw evidence, analysis inputs, and expected results are stored separately. Operation identification uses transaction execution evidence, while expected results are used for output comparisons.

\subsection{Validation Design and Results}\label{app:D.2}

Validation separately checks implementation consistency and historical episode reconstruction. Operation classifications are compared with existing results generated by the research version using the same decision rules. This checks whether organizing the analysis into a toolkit preserves its outputs. Reconstructed allowances and balances are compared with historical query values to check state recovery. Actual spending is compared with verified execution records to check callers, amounts, and permission use.

Table~\ref{tab:D1} summarizes validation results for 170 candidates and 5 historical scenarios. Validation inputs were fixed before execution, and the 5 scenarios were selected from existing research records. Counts refer to candidate events, relationship windows, transactions, or rule decisions, as applicable.

\begin{table}[!htbp]
\centering\footnotesize
\renewcommand{\thetable}{D1}
\setlength{\tabcolsep}{4pt}
\renewcommand{\arraystretch}{1.16}
\caption{Toolkit Analysis Tasks, Validation References, and Results}\label{tab:D1}
\begin{tabular}{>{\raggedright\arraybackslash}p{\dimexpr 0.200000\textwidth-2\tabcolsep\relax}>{\raggedright\arraybackslash}p{\dimexpr 0.450000\textwidth-2\tabcolsep\relax}>{\raggedright\arraybackslash}p{\dimexpr 0.350000\textwidth-2\tabcolsep\relax}}
\toprule
\textbf{Validation task} & \textbf{Inputs and comparison reference} & \textbf{Result} \\
\midrule
Operation classification & 170 positive allowance log candidates; results fixed in advance from the research program & 170 matches: 152 settings and 18 consumption events \\
Transaction-to-evidence mapping & 170 sets of transactions and receipts from two sources; call traces for 20 complex candidates (9 transactions) & All mapping checks for transaction identities, complete logs, and call traces passed \\
State reconstruction & Historical observation windows for 5 relationship groups; on-chain historical state query values & Post-start balances matched in 21/21 cases; recoverable allowances matched in 9/9 cases \\
Identification of the application that actually transferred funds & Successful execution call records and verified results for 5 selected transactions & Callers, amounts, and spending results matched for all 5 \\
Supplementary operation checks & Logs for Socket's subsequent positive allowance setting and Dexible's allowance consumption to zero & 2 matches; consumption to zero classified as consumption \\
Monitoring rules & 5 sets of verified states and remediation scopes, each evaluated under 5 rules & 25 matching decisions \\
Recomputation of specified results & 403 design-set relationships and 21 external relationships, with corresponding operations and fund-flow episodes & Zero-setting, state, episode-grouping, and rule summaries all matched \\
\bottomrule
\end{tabular}
\end{table}

The 5 scenarios include Unibot U2, a positive allowance setting after a zero allowance in Socket, and a retained allowance after Socket's old entry-point closure. They also include Multichain's same-block inflow and spending, and the Dexible control comparison. The first four reconstruct documented fund-flow episodes. For Dexible, checks separately examine allowance consumption in the reference incident transaction and balance changes within the formal observation window. Coverage of the complete account operation history at the reference block is recorded as unknown.

Allowance recovery follows each token's historical implementation. UNIBOT and TRU consumption records allow recovery of the corresponding allowance changes. Changes in USDC and WETH without direct allowance records remain unknown, while exact historical query values provide verified state points. Validation accordingly checks states that can actually be reconstructed, along with the corresponding calls and rule execution results.

Transaction identities, complete receipt logs, and necessary call traces are matched for each transaction, with failed and unknown records retained in their original denominators. Validation with real inputs and recomputation of study summaries were completed on Windows in an isolated CPython 3.13.7 virtual environment. Of 140 tests, 135 passed and 5 were skipped because of local symbolic-link permission restrictions. All 39 additional targeted tests passed.

\subsection{Code Version, Supporting Evidence, and Reproduction Procedure}\label{app:D.3}

The toolkit is version v0.1.0, released under the MIT license, and supports Python 3.10 and later, with only Python standard-library runtime dependencies. The core code and selected validation inputs used in this paper are available at \href{https://github.com/dcszhang/Authorization-Recovery-Toolkit/tree/30e9e2fe5a4423233568241b97f089484c5b52d2}{fixed code version 30e9e2f}. This version provides a consistent code and input starting point for reproduction.

Reproduction has three steps. First, validation with real inputs compares operation, state, and call results individually. Second, reviewed records are read to regenerate study summaries. Third, those summaries are compared with separately stored expected results. Outputs include individual results, sources, unknown records, and reasons for differences. The repository's Chinese and English instructions provide installation procedures, commands, and input formats; execution takes place locally and offline.

The public toolkit includes code, selected transaction evidence, examples labeled as synthetic data, and materials for recomputing specified study results. Accompanying research evidence is organized by the topics in Table~\ref{tab:D2} and preserves complete relationship identifiers, transaction locations, weights, and source mappings. Complete historical code reviews, interval estimates, and figure source data are included in the research evidence materials.

\begin{table}[!htbp]
\centering\footnotesize
\renewcommand{\thetable}{D2}
\setlength{\tabcolsep}{4pt}
\renewcommand{\arraystretch}{1.16}
\caption{Index of Research Evidence Maintained Separately from the Preprint Source}\label{tab:D2}
\begin{tabular}{>{\raggedright\arraybackslash}p{\dimexpr 0.200000\textwidth-2\tabcolsep\relax}>{\raggedright\arraybackslash}p{\dimexpr 0.310000\textwidth-2\tabcolsep\relax}>{\raggedright\arraybackslash}p{\dimexpr 0.490000\textwidth-2\tabcolsep\relax}}
\toprule
\textbf{Verification topic} & \textbf{Material} & \textbf{Main contents} \\
\midrule
Cases and remediation & Case evidence index & Contract identities, times, historical blocks, and source mappings \\
Selected fund-flow episodes & Selected-episode transaction list & Complete relationship identifiers, transaction and block hashes, UTC times, and raw amounts \\
Sampling and weighting & Sampling configurations and weights & Strata, sample sizes, weights, and related fields \\
Sources of permission at the end of the window & Day-30 permission history list & Relationship states, permission operations, and coverage \\
Unknown records and subgroup estimates & Sensitivity estimates table & Numerators, denominators, weighted estimates, and 95\% intervals \\
Time boundaries and filtering conditions & Time-boundary and sensitivity counts table & Alternative time anchors and individual counts under each condition \\
Case composition & Protocol-composition sensitivity table & Descriptive results with all cases and with each case removed \\
\bottomrule
\end{tabular}
\end{table}

\section{Remediation Scope and Timing Analysis for External Cases}\label{app:E}

\subsection{Study Objects and Actual Scope of Control}\label{app:E.1}

External validation examines whether the frozen analysis rules apply to other protocols. Dexible is studied with TRU and Multichain with WETH, both on Ethereum mainnet. Dexible's frame contains 1 relationship, which is included in full. Multichain includes 20 selected relationships, with weights of 513/4 for the earlier-activity stratum and 4060/16 for the recently active stratum.

\begin{table}[!htbp]
\centering\footnotesize
\renewcommand{\thetable}{E1}
\setlength{\tabcolsep}{4pt}
\renewcommand{\arraystretch}{1.16}
\caption{Study Objects and Disclosure Times for External Cases. Disclosure intervals are left-closed and right-open; the main analysis starts at the interval's upper bound.}\label{tab:E1}
\begin{tabular}{>{\raggedright\arraybackslash}p{\dimexpr 0.120000\textwidth-2\tabcolsep\relax}>{\raggedright\arraybackslash}p{\dimexpr 0.080000\textwidth-2\tabcolsep\relax}>{\raggedright\arraybackslash}p{\dimexpr 0.290000\textwidth-2\tabcolsep\relax}>{\raggedright\arraybackslash}p{\dimexpr 0.260000\textwidth-2\tabcolsep\relax}>{\raggedright\arraybackslash}p{\dimexpr 0.250000\textwidth-2\tabcolsep\relax}}
\toprule
\textbf{Case} & \textbf{Token} & \textbf{Approved application address} & \textbf{Disclosure interval, UTC} & \textbf{Observation start and remediation} \\
\midrule
Dexible & TRU & \texttt{0xde62e1\allowbreak{}b0edaa55\allowbreak{}aac5ffbe\allowbreak{}21984d32\allowbreak{}17064180\allowbreak{}24} & [2023-02-17 08:05:30.567, 08:05:30.568) & Interval upper bound; the original affected entry point was paused at 04:59:23 \\
Multichain & WETH & \texttt{0x6b7a87\allowbreak{}899490ec\allowbreak{}e95443e9\allowbreak{}79ca9485\allowbreak{}cbe7e715\allowbreak{}22} & [2022-01-16 10:00, 2022-01-19 12:00) & Upper bound for the main analysis, lower bound for sensitivity analysis; closure of the old entry point is unknown \\
\bottomrule
\end{tabular}
\end{table}

The TRU token contract address is \texttt{0x4c1959\allowbreak{}6f5aaff4\allowbreak{}59fa38b0\allowbreak{}f7ed92f1\allowbreak{}1ae65437\allowbreak{}84}; the WETH address is the same as in the Sushi case in Table~\ref{tab:A2}.

Dexible paused the original affected entry point at 04:59:23 UTC on 2023-02-17, before the fixed public alert \cite{ref16}. The project later also reported the pause retrospectively \cite{ref17}. The pause check in historical code, the recorded pause-state change from 0 to 1, and complete control events jointly support continued closure through day 30. The pause state and contract implementation remained unchanged throughout the observation window. The reference incident transaction contains an approval log recording allowance consumption to zero; there were zero new approval logs within the formal window.

Multichain's pool-system fix and its request for users to revoke approvals granted to old contracts address different objects \cite{ref05,ref06,ref18}. This study tracks an old contract whose permissions still require account-level cleanup; closure of its original affected entry point is recorded as unknown. This distinction determines which control evidence informs monitoring of each approval relationship.

\subsection{How Results Change When Disclosure Timing Is Uncertain}\label{app:E.2}

Multichain's disclosure time is bounded by an interval. The main analysis starts at the interval's end, while the sensitivity analysis observes 30 days from its start. Both analyses use the same 20 relationships, with each time point mapped to the exact historical block.

\begin{table}[!htbp]
\centering\footnotesize
\renewcommand{\thetable}{E2}
\setlength{\tabcolsep}{4pt}
\renewcommand{\arraystretch}{1.16}
\caption{Multichain Results with Two Observation Starts. D denotes the selected disclosure start, and positive A means that the allowance remains greater than zero. States are counted by relationship and returning funds by episode.}\label{tab:E2}
\begin{tabular}{>{\raggedright\arraybackslash}p{\dimexpr 0.230000\textwidth-2\tabcolsep\relax}>{\raggedright\arraybackslash}p{\dimexpr 0.115000\textwidth-2\tabcolsep\relax}>{\raggedright\arraybackslash}p{\dimexpr 0.115000\textwidth-2\tabcolsep\relax}>{\raggedright\arraybackslash}p{\dimexpr 0.115000\textwidth-2\tabcolsep\relax}>{\raggedright\arraybackslash}p{\dimexpr 0.115000\textwidth-2\tabcolsep\relax}>{\raggedright\arraybackslash}p{\dimexpr 0.170000\textwidth-2\tabcolsep\relax}>{\raggedright\arraybackslash}p{\dimexpr 0.140000\textwidth-2\tabcolsep\relax}}
\toprule
\textbf{Observation start} & \textbf{Positive allowance at D} & \textbf{Positive allowance at day 1} & \textbf{Positive allowance at day 7} & \textbf{Positive allowance at day 30} & \textbf{Episodes of returning funds} & \textbf{Of these, use of a pre-disclosure allowance} \\
\midrule
Interval end: main analysis & 17 & 17 & 14 & 11 & 3 & 1 \\
Interval start: sensitivity analysis & 19 & 19 & 16 & 11 & 2 & 1 \\
\bottomrule
\end{tabular}
\end{table}

Before the disclosure interval began, the fixed initial group contained 19 relationships with positive allowances. Of these, 2 had their allowances set to zero within the interval, leaving 17 positive allowances at the main-analysis start. These 2 remain in the denominator of the original 19 relationships. In total, 8 relationships in the initial group had allowances explicitly set to zero: 6 within the main window and 2 within the disclosure interval.

Both zero-setting calculations use the same weights. The initially positive allowance group's weights sum to 17277/4; those of the 6 relationships set to zero within the main window sum to 3045/2. Including the 2 relationships set to zero within the interval raises the zero-setting weight sum to 2030. Thus, the main-window zero-setting proportion is approximately 35.2\%, and the cumulative proportion including interval events is approximately 47.0\%. Under both timing definitions, zero-setting progress remains below 50\% and 90\% through day 30, so \(t_{50}\) and \(t_{90}\) are recorded as ``not reached within the window.''

Starting at the interval's beginning also moves the end of the 30-day window earlier. One inflow to a relationship with a zero allowance occurs after that window ends, reducing episodes of returning funds from 3 in the main analysis to 2. Same-block use of a pre-disclosure allowance falls within both observation windows, and 11 relationships retain allowances at day 30 under either starting point.

\subsection{Transaction Order for Same-Block Inflow and Use of a Pre-Disclosure Allowance}\label{app:E.3}

Table~\ref{tab:E3} details a Multichain episode in which 1.4 WETH entered an account and was immediately transferred out using its pre-disclosure allowance. The inflow and spending occurred in adjacent transactions within the same block, with a zero account balance both before and after block execution. Expanding the transaction sequence reveals the intervening inflow and permission use.

\begin{table}[!htbp]
\centering\footnotesize
\renewcommand{\thetable}{E3}
\setlength{\tabcolsep}{4pt}
\renewcommand{\arraystretch}{1.16}
\caption{States and Transaction Order in the Selected Multichain Episode. WETH uses 18 decimal places, so the raw integer 1400000000000000000 corresponds to 1.4 WETH; MAX denotes the maximum allowance. ``\#'' after a block number indicates the transaction index.}\label{tab:E3}
\begin{tabular}{>{\raggedright\arraybackslash}p{\dimexpr 0.150000\textwidth-2\tabcolsep\relax}>{\raggedright\arraybackslash}p{\dimexpr 0.160000\textwidth-2\tabcolsep\relax}>{\raggedright\arraybackslash}p{\dimexpr 0.120000\textwidth-2\tabcolsep\relax}>{\raggedright\arraybackslash}p{\dimexpr 0.220000\textwidth-2\tabcolsep\relax}>{\raggedright\arraybackslash}p{\dimexpr 0.120000\textwidth-2\tabcolsep\relax}>{\raggedright\arraybackslash}p{\dimexpr 0.230000\textwidth-2\tabcolsep\relax}}
\toprule
\textbf{Stage} & \textbf{Block and transaction position} & \textbf{Balance (WETH)} & \textbf{Raw integer balance} & \textbf{Allowance} & \textbf{Verification evidence} \\
\midrule
Before inflow & End of 14121705 & 0 & 0 & MAX & States agree across two independent sources \\
Incoming funds & 14121706 \#0 & 1.4 & 1400000000000000000 & MAX & \href{https://etherscan.io/tx/0xddded9ebde98b6dec30d2ed5bbb89263a743642f1b488078acfb160a1d6b6486}{Incoming transaction} \\
Use of a pre-disclosure allowance & 14121706 \#1 & 0 & 0 & MAX & \href{https://etherscan.io/tx/0x72930a51a39ee7fd62b1570bbcca303e1b6a3734c338542554276eb735864d09}{Spending transaction} \\
Allowance explicitly set to zero & 14121735 & 0 & 0 & 0 & \href{https://etherscan.io/tx/0x0052afe8c3c9beb0745ceb66259aedcf9cc59d84dc1689aeb6be911101d1840e}{Zero-setting transaction} \\
\bottomrule
\end{tabular}
\end{table}

The account address for this relationship is \texttt{0xf94b11\allowbreak{}be5fb9fc\allowbreak{}bcf90950\allowbreak{}4d313f1c\allowbreak{}b8a3075d\allowbreak{}22}. The block containing the inflow and spending is timestamped 2022-02-01 at 18:26:15 UTC. The allowance was set to zero at 18:31:50, an interval of 335 seconds.

Execution records confirm that the old contract in this case actually called the token contract to transfer funds. The call passed through the same affected entry point as the reference incident. No positive allowance-setting operation occurred between disclosure and this spending, so the spending used the pre-disclosure allowance. WETH retains the allowance when MAX is used; this permission therefore persisted after the funds left, until the later zero-setting operation.

Historical code, calls, and before-and-after states from independent sources jointly support this episode. Its exact sequence can be reviewed through the transactions in Table~\ref{tab:E3}. The transfer direction is labeled third-party-directed, and its specific purpose is recorded as unknown.

\end{document}